\documentclass{new_tlp}

\usepackage[utf8]{inputenc}
\usepackage{microtype}
\usepackage{xparse,xifthen}
\usepackage{natbib}
\usepackage{amsmath,amssymb}
\let\proof\relax

\usepackage{amsthm}
\usepackage{url}
\usepackage{listings}
\usepackage[dvipsnames]{xcolor}
\usepackage{tikz}
\usepackage{slashbox}

\makeatletter
\let\O@argtabularcr\@argtabularcr
\def\O@xtabularcr{\@ifnextchar[\O@argtabularcr{\ifnum 0=`{\fi}\cr}}
\let\O@tabacol\@tabacol
\let\O@tabclassiv\@tabclassiv
\let\O@tabclassz\@tabclassz
\let\O@tabarray\@tabarray
\def\author@tabular{\authorsize\def\@halignto{}\@authortable}
\let\endauthor@tabular=\endtabular
\def\author@tabcrone{{\ifnum0=`}\fi\O@xtabularcr\affilsize\itshape
 \let\\=\author@tabcrtwo\ignorespaces}
\def\author@tabcrtwo{{\ifnum0=`}\fi\O@xtabularcr[-3\p@]\affilsize\itshape
 \let\\=\author@tabcrtwo\ignorespaces}
\def\@authortable{\leavevmode \hbox \bgroup $\let\@acol\O@tabacol
 \let\@classz\O@tabclassz \let\@classiv\O@tabclassiv
 \let\\=\author@tabcrone \ignorespaces \O@tabarray}
\makeatother

\usepackage{xcolor,colortbl}
\usepackage{xintexpr, xinttools}
\usepackage{caption}

\newcommand{\review}[1]{{\textcolor{black}{#1}}}

\newcommand{\Reactants}[1]{\ensuremath{\mathit{rcts}(#1)}} 
\newcommand{\Products}[1]{\ensuremath{\mathit{prds}(#1)}} 

\newcommand{\StoichiometricFunction}{\ensuremath{s}} 
\newcommand{\Stoichiometry}[2]{\ensuremath{\StoichiometricFunction(#1,#2)}}
\newcommand{\ExportReaction}{\ensuremath{r_e}}
\newcommand{\SeedReaction}{\ensuremath{r_{s_{1}}}}
\newcommand{\SeedReactiont}{\ensuremath{r_{s_{2}}}}

\newcommand{\ActivityFunction}{\ensuremath{\mathit{active}}}
\newcommand{\Activity}[3]{\ensuremath{\ActivityFunction^{#1}_{#2}(#3)}}
\newcommand{\Activitytwo}[2]{\ensuremath{\ActivityFunction_{#1}(#2)}}
\newcommand{\bb}{\textsc{bb}}
\newcommand{\usc}{\textsc{usc}}
\newcommand{\opts}{\textsc{\#opts}}
\newcommand{\sols}{\textsc{\#sols}}
\newcommand{\f}{\textsc{f}}

\newcommand{\verified}{\textsc{verified}}
\newcommand{\vbs}{\textsc{vbs}}
\newcommand{\T}{\textsc{t}}
\newcommand{\TO}{\textsc{to}}

\newcommand{\degradation}{\textsc{degradation}}
\newcommand{\prop}[1]{\textsc{prop}-#1}
\newcommand{\core}[1]{\textsc{core}-#1}

\newcommand{\default}{\textsc{default}}

\newcommand{\strseed}{S_{b}\,}

\newcommand{\MinFlux}[1]{\ensuremath{\mathit{lb}_#1}}
\newcommand{\MaxFlux}[1]{\ensuremath{\mathit{ub}_#1}}

\newcommand{\sysfont}{\textit}

\newcommand{\clingo}{\sysfont{clingo}}
\newcommand{\cplex}{\sysfont{cplex}}

\newcommand{\cobrapy}{\sysfont{cobrapy}}

\newcommand{\fluto}{\sysfont{fluto}}

\newcommand{\gapfill}{\sysfont{gapfill}}

\newcommand{\lpsolve}{\sysfont{lpsolve}}

\newcommand{\meneco}{\sysfont{meneco}}

\newcommand{\python}{Python}

\newtheorem{theorem}{Theorem}

\allowdisplaybreaks

\title{Hybrid Metabolic Network Completion}

\author[C.~Frioux, T.~Schaub, S.~Schellhorn, A.~Siegel, and P.~Wanko]{%
  Clémence Frioux
  \\
  Univ Rennes, Inria, CNRS, IRISA F-35000 Rennes, France
  \and
  Torsten Schaub
  \\
  Inria, Rennes, France \ and \ Universit\"at Potsdam, Germany
  \and
  Sebastian Schellhorn
  \\
  Universit\"at Potsdam, Germany
  \and
  Anne Siegel
  \\
  Univ Rennes, Inria, CNRS, IRISA F-35000 Rennes, France
  \and
  Philipp Wanko
  \\
  Universit\"at Potsdam, Germany} %

\begin{document}

\maketitle
\begin{abstract}
Metabolic networks play a crucial role in biology since they capture all chemical reactions in an organism.
While there are networks of high quality for many model organisms,
networks for less studied organisms are often of poor quality and suffer from incompleteness.
To this end, we introduced in previous work an ASP-based approach to metabolic network completion.
Although this qualitative approach allows for restoring moderately degraded networks,
it fails to restore highly degraded ones.
This is because it ignores quantitative constraints capturing reaction rates.
To address this problem, we propose a hybrid approach to metabolic network completion
that integrates our qualitative ASP approach with quantitative means for capturing reaction rates.
We begin by formally reconciling existing stoichiometric and topological approaches to network completion in a unified formalism.
With it, we develop a hybrid ASP encoding and rely upon the theory reasoning capacities of the ASP system \clingo\
for solving the resulting logic program with linear constraints over reals.
We empirically evaluate our approach by means of the metabolic network of \emph{Escherichia coli}.
Our analysis shows that our novel approach yields greatly superior results than obtainable from purely qualitative or quantitative approaches.
\textit{Under consideration in Theory and Practice of Logic Programming (TPLP).}
\end{abstract}



\section{Introduction}\label{sec:introduction}
\newcommand\blfootnote[1]{%
  \begingroup
  \renewcommand\thefootnote{}\footnote{#1}%
  \addtocounter{footnote}{-1}%
  \endgroup
}
\blfootnote{This is an extended version of a paper presented at LPNMR-17, invited as a rapid publication in TPLP. The authors acknowledge the assistance of the conference chairs Tomi Janhunen and Marco Balduccini.}
Among all biological processes occurring in a cell, metabolic networks are in charge of transforming
input nutrients into both energy and output nutrients necessary for the functioning of other cells.
In other words, they capture all chemical reactions occurring in an organism.
In biology,
such networks are crucial from a fundamental and technological point of view
to estimate and control the capability of organisms to produce certain products.
Metabolic networks of high quality exist for many model organisms.
In addition,
recent technological advances enable their semi-automatic generation for many less studied organisms, also described as non-model organisms.
However,
the resulting metabolic networks are usually of poor quality,
due to error-prone, genome-based construction processes and a lack of (human) resources.
As a consequence, they usually suffer from substantial incompleteness.
The common fix is to fill the gaps by completing a draft network by borrowing chemical pathways
from reference networks of well studied organisms until the augmented network provides the measured functionality.

In previous work~\citep{schthi09a}, we introduced a logical approach to \emph{metabolic network completion}
by drawing on the work in~\citep{haebhe05a}. 
We formulated the problem as a qualitative combinatorial (optimization) problem and solved it with Answer Set Programming (ASP~\citep{baral02a}).
The basic idea is that reactions apply only if all their reactants are available,
either as nutrients or provided by other metabolic reactions.
Starting from given nutrients, referred to as \emph{seeds},
this allows for extending a metabolic network by successively adding operable
reactions and their products.
The set of compounds in the resulting network is called the \emph{scope} of the
seeds and represents all compounds that can principally be synthesized from the seeds.
In metabolic network completion, we query a database of metabolic reactions
looking for (minimal) sets of reactions that can restore an observed bio-synthetic behavior.
This is usually expressed by requiring that certain \emph{target} compounds are in the scope of some given seeds.
For instance, in the follow-up work in~\citep{coevgeprscsith13a,prcodideetdaevthcabosito14a},
we successfully applied our ASP-based approach to the reconstruction of the metabolic network of the macro-algae \emph{Ectocarpus siliculosus},
using the collection of \review{reference networks Metacyc \citep{Caspi2016}.}

We evidenced in~\citep{Prigent2017}
that our ASP-based method partly restores the bio-synthetic capabilities of a large proportion of moderately degraded networks: it fails to restore the ones of both some moderately degraded and most of highly degraded metabolic networks.
The main reason for this is that our purely qualitative approach misses quantitative constraints
accounting for the law of mass conservation,
a major hypothesis about metabolic networks.
This law stipulates that
each internal metabolite of a network must balance its production rate with its consumption rate at the steady state of the system.
Such rates are given by the weighted sums of all reaction rates consuming or producing a metabolite, respectively.
This calculation is captured by the \emph{stoichiometry}\footnote{See also \url{https://en.wikipedia.org/wiki/Stoichiometry}.} of the involved reactions.
Hence,
the qualitative ASP-based approach fails to tell apart solution candidates with correct and incorrect stoichiometry
and therefore reports inaccurate results for some degraded networks.

We address this by proposing a hybrid approach to metabolic network completion that integrates our qualitative ASP approach
with quantitative techniques from
\emph{Flux Balance Analysis} (FBA\footnote{See also \url{https://en.wikipedia.org/wiki/Flux_balance_analysis}.}~\citep{marzom16a}), 
the state-of-the-art quantitative approach for capturing reaction rates in metabolic networks.
We accomplish this by taking advantage of recently developed theory reasoning capacities for the ASP system \clingo~\citep{gekakaosscwa16a}.
More precisely,
we use an extension of \clingo\ with linear constraints over reals, as dealt with in Linear Programming (LP~\citep{dantzig63a}).
This extension provides us with an extended ASP modeling language as well as a generic interface to alternative LP solvers, viz.\ \cplex\ and \lpsolve,
for dealing with linear constraints.
We empirically evaluate our approach by means of the metabolic network of \emph{Escherichia coli}.
Our analysis shows that our novel approach yields superior results than obtainable from purely qualitative or quantitative approaches.
Moreover, our hybrid application provides a first evaluation of the theory extensions of the ASP system \clingo\
with linear constraints over reals in a non-trivial setting.
%



\section{Metabolic Network Completion}\label{sec:problem}

Metabolism is the sum of all chemical reactions occurring within an organism.
As the products of a reaction may be reused as reactants, reactions can be chained to complex chemical pathways.
Such complex pathways are described by a metabolic network.

We represent a \emph{metabolic network} as a labeled directed bipartite graph
\(
G=(R\cup M,E,\StoichiometricFunction),
\)
where $R$ and $M$ are sets of nodes standing for \emph{reactions} and \emph{compounds} (also called metabolites), respectively.
When $(m,r)\in E$ or $(r,m)\in E$ for $m\in M$ and $r\in R$, the metabolite $m$ is called a \emph{reactant} or \emph{product} of reaction~$r$, respectively. \review{Metabolites and reactions nodes can both have multiple ingoing and outgoing edges.}
More formally, for any $r\in R$, define
\(
\Reactants{r}=\{m\in M\mid (m,r)\in E\}
\)
and
\(
\Products{r} =\{m\in M\mid (r,m)\in E\}
\).
The \emph{edge labeling}
\(
\StoichiometricFunction: E\rightarrow \mathbb{R}
\)
gives the stoichiometric coefficients of a reaction's reactants and products, respectively, i.e., their relative quantities involved in the reaction.
Finally, the activity rate of reactions is bound by lower and upper bounds,
denoted by $\MinFlux{r}\in\mathbb{R}^+_0$ and $\MaxFlux{r}\in\mathbb{R}^+_0$ for $r\in R$, respectively.
Whenever clear from the context,
we refer to metabolic networks with $G$ (or $G'$, etc) and denote the associated reactions and compounds with
$M$ and $R$ (or $M',R'$ etc), respectively.

We distinguish a set $S \subseteq M$ of compounds as initiation \emph{seeds}, that is,
compounds initially present due to experimental evidence.
Another set of compounds is assumed to be activated by default.
These \emph{boundary compounds} are defined as:
\(
\strseed(G) = \{ m\in M \mid r\in R,  m\in \Products{r}, \Reactants{r}=\emptyset \}
\).
For simplicity, we assume that all boundary compounds are seeds: $\strseed(G)\subseteq S$.
Note that follow-up concepts like reachability and activity in network completion are independent of this assumption.

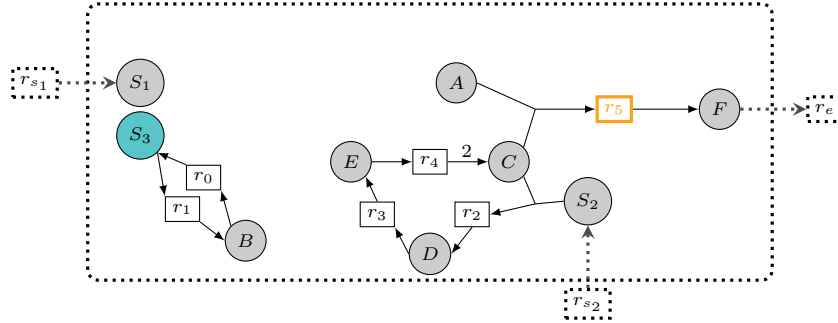
\begin{figure}[t]
  \centering
\usetikzlibrary{shapes.misc, positioning}
\begin{tikzpicture}[scale=0.7]\scriptsize
  \tikzstyle{metabolite}=[draw,circle,fill=white!80!black];
  \tikzstyle{repairmetabolite}=[draw,white!40!black, circle,fill=white!90!black,text=white!40!black,dashed];
  \tikzstyle{seed}=[draw,circle,fill=BlueGreen!70];
  \tikzstyle{target}=[draw,circle,fill=YellowOrange];
  \tikzstyle{reaction}=[draw,rectangle];
   \tikzstyle{export}=[draw,rectangle,dotted, very thick];
   \tikzstyle{exportrepair}=[draw,rectangle,dotted, very thick,white!80!black,text=white!70!black];
  \tikzstyle{repairreaction}=[draw,rectangle,white!40!black,text=white!40!black,dashed];
  \tikzstyle{initial}=[->,>=latex,thick];
  \tikzstyle{bdd}=[->,>=latex,thick];
  \tikzstyle{etiq}=[midway,fill=black!20,scale=0.5];
  \tikzstyle{stc}=[draw, rectangle, white, text=black]

  \node[stc] (stcr4C) at (6.2,6.7) {$2$};

  \draw [black,dotted, rounded corners, very thick] (-1,4.25) rectangle (12,9.5);

  \node[seed] (S2) at (0,7) {$S_{3}$};
  \node[metabolite] (Sb2) at (8.5,5.75) {$S_{2}$};
  \node[metabolite] (S1) at (0,8) {$S_{1}$};

  \node[metabolite] (F) at (11,7.50) {$F$};

  \node[metabolite] (A) at (6,8) {$A$};
  \node[metabolite] (B) at (2,5.0) {$B$};
  \node[metabolite] (C) at (7,6.5) {$C$};
  \node[metabolite] (D) at (5.50,4.75) {$D$};
  \node[metabolite] (E) at (4,6.5) {$E$};

  \node[reaction] (R0) at (1.2,6.2) {$r_{0}$};
  \node[reaction] (R1) at (0.8,5.60) {$r_{1}$};
  \node[reaction] (R2) at (6.3,5.5) {$r_{2}$};
  \node[reaction] (R3) at (4.5,5.5) {$r_{3}$};
  \node[reaction] (R4) at (5.5,6.5) {$r_{4}$};
  \node[reaction, very thick,YellowOrange] (R5) at (9,7.50) {$r_{5}$}; 

  \draw[->,>=latex] (B.north west) -- (R0.south east);
  \draw[->,>=latex] (R0.north west) -- (S2.south east);

  \draw[->,>=latex] (S2.south east) -- (R1.north west);
  \draw[->,>=latex] (R1.south east) -- (B.west);

  \draw[->,>=latex] (Sb2.west) -- (7.5,5.70) -- (R2.east);
  \draw[] (C.south east) -- (7.5,5.70);
  \draw[->,>=latex] (R2.south) -- (D.east);

  \draw[->,>=latex] (D.west) -- (R3.south east);
  \draw[->,>=latex] (R3.north) -- (E.south east);

  \draw[->,>=latex] (E.east) -- (R4.west);
  \draw[->,>=latex] (R4.east) -- (C.west);

  \draw[->,>=latex] (A.east) -- (7.5,7.50) -- (R5.west);
  \draw[] (C.north east) -- (7.5,7.50);
  \draw[->,>=latex] (R5.east) -- (F.west);






  \node[export] (outF) at (13,7.5) {\ExportReaction};
   \draw[->,>=stealth,white!30!black,dotted, very thick] (F.east) --  (outF.west);

   \node[export] (inS) at (-2,8) {$r_{s_1}$};
   \draw[->,>=stealth,white!30!black,dotted, very thick] (inS) --  (S1.west);

    \node[export] (inS2) at (8.5,3.8) {$r_{s_2}$};
    \draw[->,>=stealth,white!30!black,dotted, very thick] (inS2) --  (Sb2.south);

\end{tikzpicture}
%
    \caption{Example of a metabolic network. Compounds and reactions are depicted by circles and rectangles respectively. Dashed reactions are reactions involving the boundary between the organism's metabolism and its environment. $r_5$ is the target reaction. $S_1$ and $S_2$ are boundary (and initiation) seeds. $S_3$ is assumed to be an initiation seed. Numbers on arrows describe the stoichiometry of reaction (default value is 1).}
    \label{gra:toy_d}
\end{figure}

For illustration, consider the metabolic network in Fig.~\ref{gra:toy_d}.
%
%
The network consists of 9 reactions, \SeedReaction, \SeedReactiont, \ExportReaction{} and $r_0$ to $r_5$, and 8 compounds, $A,\dots,F$, $S_1$, $S_2$ and $S_3$.
Here, $S=\{S_1,S_2,S_3\}$, $S_1$ and $S_2$ being the two boundary compounds of the network. Dashed rectangle describes the boundary of the system, outside of which is the environment of the organism.
Consider reaction
\(
r_4 : E\rightarrow 2C
\)
transforming one unit of $E$ into two units of $C$ (stoichiometric coefficients of 1 are omitted in the graphical representation; cf.~Fig.~\ref{gra:toy_d}).
We have
$\Reactants{r_4}=\{E\}$,
$\Products{r_4}=\{C\}$,
along with $\Stoichiometry{E}{r_4}=1 $
\ and $\Stoichiometry{r_4}{C}=2$.

In biology, several concepts have been introduced to model the activation of reaction fluxes in metabolic networks,
or to synthesize metabolic compounds.
To model this,
we introduce a function \ActivityFunction\ that given a metabolic network $G$ takes a set of seeds $S \subseteq M$ and returns a set of activated
reactions $\Activitytwo{G}{S} \subseteq R$.
%
%
\begin{figure}[t]
  \centering
\usetikzlibrary{shapes.misc, positioning}
\begin{tikzpicture}[scale=0.7]\scriptsize
  \tikzstyle{metabolite}=[draw,circle,fill=white!80!black];
  \tikzstyle{repairmetabolite}=[draw,white!40!black, circle,fill=white!90!black,text=white!40!black,dashed];
  \tikzstyle{seed}=[draw,circle,fill=BlueGreen!70];
  \tikzstyle{target}=[draw,circle,fill=YellowOrange];
  \tikzstyle{reaction}=[draw,rectangle];
   \tikzstyle{export}=[draw,rectangle,dotted, very thick];
   \tikzstyle{exportrepair}=[draw,rectangle,dotted, very thick,white!80!black,text=white!70!black];
  \tikzstyle{repairreaction}=[draw,rectangle,white!40!black,text=white!40!black,dashed];
  \tikzstyle{initial}=[->,>=latex,thick];
  \tikzstyle{bdd}=[->,>=latex,thick];
  \tikzstyle{etiq}=[midway,fill=black!20,scale=0.5];
  \tikzstyle{stc}=[draw, rectangle, white, text=black]

  \node[stc] (stcr4C) at (6.2,6.7) {$2$};

  \draw [black,dotted, rounded corners, very thick] (-1,4.25) rectangle (12,9.5);

  \node[seed] (S2) at (0,7) {$S_{3}$};
  \node[metabolite] (Sb2) at (8.5,5.75) {$S_{2}$};
  \node[metabolite] (S1) at (0,8) {$S_{1}$};

  \node[metabolite] (F) at (11,7.50) {$F$};

  \node[metabolite] (A) at (6,8) {$A$};
  \node[metabolite] (B) at (2,5.0) {$B$};
  \node[metabolite] (C) at (7,6.5) {$C$};
  \node[metabolite] (D) at (5.50,4.75) {$D$};
  \node[metabolite] (E) at (4,6.5) {$E$};
  \node[repairmetabolite] (X) at (5,9) {$G$};

  \node[reaction] (R0) at (1.2,6.2) {$r_{0}$};
  \node[reaction] (R1) at (0.8,5.60) {$r_{1}$};
  \node[reaction] (R2) at (6.3,5.5) {$r_{2}$};
  \node[reaction] (R3) at (4.5,5.5) {$r_{3}$};
  \node[reaction] (R4) at (5.5,6.5) {$r_{4}$};
  \node[reaction, very thick,YellowOrange] (R5) at (9,7.50) {$r_{5}$}; 
  \node[repairreaction] (R6) at (3,8) {$r_{6}$};
  \node[repairreaction] (R7) at (2.3,6.9) {$r_{7}$};
  \node[repairreaction] (R8) at (3,5.8) {$r_{8}$};
  \node[repairreaction] (R9) at (8.5,8.5) {$r_{9}$};

  \draw[->,>=latex] (B.north west) -- (R0.south east);
  \draw[->,>=latex] (R0.north west) -- (S2.south east);

  \draw[->,>=latex] (S2.south east) -- (R1.north west);
  \draw[->,>=latex] (R1.south east) -- (B.west);

  \draw[->,>=latex] (Sb2.west) -- (7.5,5.70) -- (R2.east);
  \draw[] (C.south east) -- (7.5,5.70);
  \draw[->,>=latex] (R2.south) -- (D.east);

  \draw[->,>=latex] (D.west) -- (R3.south east);
  \draw[->,>=latex] (R3.north) -- (E.south east);

  \draw[->,>=latex] (E.east) -- (R4.west);
  \draw[->,>=latex] (R4.east) -- (C.west);

  \draw[->,>=latex] (A.east) -- (7.5,7.50) -- (R5.west);
  \draw[] (C.north east) -- (7.5,7.50);
  \draw[->,>=latex] (R5.east) -- (F.west);

  \draw[->,>=latex,white!55!black,dashed] (S1.east) -- (R6.west);
  \draw[->,>=latex,white!55!black,dashed] (R6.east) -- (4,8) -- (A.west);
  \draw[->,>=latex,white!55!black,dashed] (4,8) -- (X.west);

  \draw[->,>=latex,white!55!black,dashed] (S2.east) -- (R7.west);
  \draw[->,>=latex,white!55!black,dashed] (R7.east) -- (E.north west);

  \draw[->,>=latex,white!55!black,dashed] (B.north east) -- (R8.south west);
  \draw[->,>=latex,white!55!black,dashed] (R8.east) -- (E.south);

  \draw[->,>=latex,white!55!black,dashed] (X.east) -- (R9.north west);
  \draw[->,>=latex,white!55!black,dashed] (R9.east) -- (F.north west);


  \node[export] (outF) at (13,7.5) {\ExportReaction};
   \draw[->,>=stealth,white!30!black,dotted, very thick] (F.east) --  (outF.west);

    \node[export] (inS) at (-2,8) {$r_{s_1}$};
    \draw[->,>=stealth,white!30!black,dotted, very thick] (inS) --  (S1.west);

  \node[export] (inS2) at (8.5,3.8) {$r_{s_2}$};
  \draw[->,>=stealth,white!30!black,dotted, very thick] (inS2) --  (Sb2.south);

\end{tikzpicture}
%
    \caption{Metabolic network completion problem. The purpose of its solving is to select the minimal number of reactions from a database (dashed shaded reactions) such that activation of target reaction $r_5$ is restored from boundary and/or initiation seeds. There are three formalisms for activation of target reaction: stoichiometric, topological and hybrid.}
    \label{gra:toy}
\end{figure}
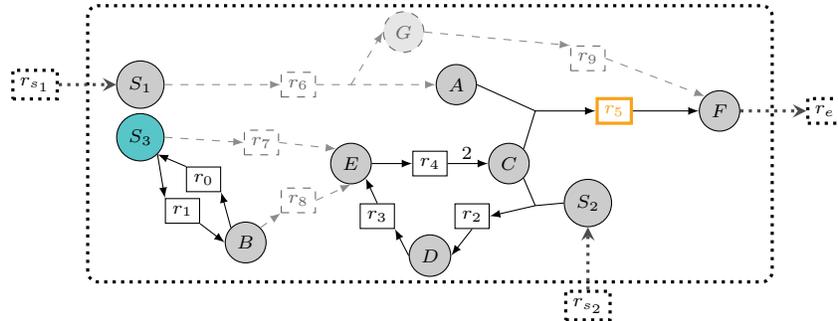
%
With it,
\emph{metabolic network completion} is about ensuring that a set of target reactions (reaction $r_5$ in~Fig.~\ref{gra:toy_d}) is activated from seed compounds in $S$
by possibly extending the metabolic network with reactions from a reference network (cf.\ shaded part in~Fig.~\ref{gra:toy}).

Formally, given
a metabolic network $G=(R\cup M,E,\StoichiometricFunction)$, 
a set $S\subseteq M$ of seed compounds such that $\strseed(G) \subseteq S$,
a set $R_{T}\subseteq R$ of target reactions, and
a reference network $(R'\cup M',E',\StoichiometricFunction')$,
the \emph{metabolic network completion problem} is to find a set $R''\subseteq R'\setminus R$ of reactions of minimal size such that
\(
R_{T}\subseteq\Activitytwo{G''}{S}
\)
where%
\footnote{Since \StoichiometricFunction, $\StoichiometricFunction'$ have disjoint domains we view them as relations and compose them by union.}
\begin{align}
  \label{eq:completion:graph}
  G''&= ((R\cup R'')\cup (M\cup M''),E\cup E'',\StoichiometricFunction'')\ ,
  \\\label{eq:completion:metabolites}
  M''&=\{m\in M'\mid r\in R'', m\in\Reactants{r}\cup\Products{r}\}\ ,
  \\\label{eq:completion:edges}
  E''&=E'\cap((M''\times R'')\cup(R''\times M'')) , \text{ and}
  \\
  \StoichiometricFunction''&=\StoichiometricFunction\cup\StoichiometricFunction'
  \ .
\end{align}
We call $R''$ a \emph{completion} of $(R\cup M,E,\StoichiometricFunction)$ from $(R'\cup M',E',\StoichiometricFunction')$ wrt $S$ and $R_{T}$.
Our concept of activation allows different biological paradigms to be captured.
Accordingly,
different formulations of metabolic network completion can be characterized:
the stoichiometric, the relaxed stoichiometric, the topological, and the hybrid one.
We elaborate upon their formal characterizations in the following sections.

\subsection{Stoichiometric Metabolic Network Completion}\label{sec:stoichio} 
The first activation semantics has been introduced in the context of Flux Balance Analysis
capturing reaction flux distributions of metabolic networks at steady state.
In this paradigm, each reaction $r$ is associated with a \emph{metabolic flux value},
expressed as a real variable $v_r$ confined by the minimum and maximum rates:
\begin{align} \label{eq:stoichiometric:bounds}
  & \MinFlux{r} \leq  v_r\leq \MaxFlux{r} \qquad\text{ for } r\in R.
\end{align}
Flux distributions are formalized in terms of a system of equations relying on the stoichiometric coefficients of reactions.
\review{Reaction stoichiometries are governed by the \emph{law of mass conservation} under a steady state assumption; in other words, the mass of the system remains constant over the reaction.
The input and output fluxes of reactions consuming and producing a metabolite are balanced.}
\begin{align}
\label{eq:stoichiometric:equation}
  & \textstyle
    \sum_{\substack{r\in R}}\Stoichiometry{r}{m}\cdot v_r
    +
    \sum_{\substack{r\in R}}-\Stoichiometry{m}{r}\cdot v_r
    =
    0
    \qquad \text{ for } m\in M.
\end{align}
Given a target reaction $r_T\in R_T$, a metabolic network $G=(R\cup M,E,\StoichiometricFunction)$ and a set of seeds $S$,
\emph{stoichiometric activation} is defined as follows:
\begin{align}
\label{eq:stoichiometric:activation}
  r_T \in \Activity{s}{G}{S} & \ \text{ iff } \ v_{r_T} >0 \text{ and }
                               \eqref{eq:stoichiometric:bounds} \text{ and } \eqref{eq:stoichiometric:equation}\text{ hold for }M\text{ and }R.
\end{align}
Note that the condition $v_{r_T} >0$ strengthens the flux condition for $r_T\in R$ in the second part.
More generally, observe that activated target reactions are not directly related to the network's seeds $S$.
However,
the activation of targets highly depends on the boundary compounds in $\strseed(G)$
for which \eqref{eq:stoichiometric:equation} \review{is always satisfied and thus initiates the fluxes.
Since boundary compounds are produced by at least one reaction without prerequisite,
an arbitrary amount might be produced.
Therefore, the incoming flux value always balances the sum of the flux values associated to outgoing edges.
Intuitively, boundary compounds are nutrients that are expected to be available in the system
for the consumption by the metabolic network,
thus initiating the reactions within.}
In our draft network $G$,
consisting of all \review{non-dashed} nodes and edges depicted in Fig.~\ref{gra:toy}
(viz.\ reactions \SeedReaction, \SeedReactiont, \ExportReaction{} and $r_0$ to $r_5$ and compounds $A,\dots,F$, $S_1$, $S_2$, and $S_3$ and $r_5$ the single target reaction)
and the reference network $G'$,
consisting of the shaded part of Fig~\ref{gra:toy},
(viz.\ reactions $r_6$ to $r_9$ and metabolite $G$)
a strict stoichiometry-based completion aims to obtain a solution with $r_5\in\Activity{s}{G''}{\{S_1,S_2,S_3\}}$ where $v_{r_5}$ is maximal.
This can be achieved by adding the completion $R''_1=\{r_{6},r_9\}$ (Fig.~\ref{gra:toy_ss}).
\review{The cycle made of compounds $E,C,D$ and the boundary seed $S_2$ is already balanced and notably self-activated.
Indeed, initiation of $D$ and $E$ producibility requires the producibility of $C$ (in addition to the presence of the boundary seed $S_2$) that itself depends on $D$ and $E$. Yet, according the flux conditions, that models steady state conditions, the cycle is activated.
Such self-activation of cyclic pathways is an inherent problem of purely stoichiometric approaches to network completion.
This is a drawback of the semantics because the effective activation of the cycle requires the additional (and unchecked) condition that at least one of the compounds was present as the initial state of the system. This could be the case provided there exist another way to enable the production of one or several components of the cycle (here an activable reaction producing $E$ for instance) \citep{Prigent2017}.}
The instance of Equation~\eqref{eq:stoichiometric:equation} controlling the reaction rates related to metabolite $C$ is
\(
2\cdot v_{r_4} - v_{r_2} - v_{r_5} = 0
\).

To solve metabolic network completion with flux-balance activated reactions,
Linear Programming can be used to maximize the flux rate $v_{r_T}$ provided that the linear constraints are satisfied.
Nonetheless, this problem turns out to be hard to solve in practice and existing approaches scale poorly to real-life applications (cf.~ \citep{Orth2010}).

This motivated the use of approximate methods.
The relaxed problem is obtained by weakening the mass-balance equation \eqref{eq:stoichiometric:equation} as follows:
\begin{align}
\label{eq:stoichiometric:equation:relaxed}
  &  \textstyle        \sum_{\substack{r\in R}} \Stoichiometry{r}{m}\cdot v_r
                       +
                       \sum_{\substack{r\in R}}-\Stoichiometry{m}{r}\cdot v_r
                       \geq
                       0
                       \qquad \text{ for } m\in M.
\end{align}
This lets us define the concept of \emph{relaxed stoichiometric activation}:
\begin{align}
\label{eq:stoichiometric:activation:relaxed}
  r_T \in \Activity{r}{G}{S} & \ \text{ iff } \ v_{r_T} >0 \text{ and }
                               \eqref{eq:stoichiometric:bounds} \text{ and } \eqref{eq:stoichiometric:equation:relaxed}\text{ hold for }M\text{ and }R.
\end{align}
The resulting problem can now be efficiently solved with Linear Programming~\citep{SatishKumar2007}.
Existing systems addressing strict stoichiometric network completion either
cannot guarantee optimal solutions~\citep{laten2014a} or
do not support a focus on specific target reactions~\citep{Thiele2014}.
Other approaches either partially relax the problem~\citep{Vitkin2012} or
solve the relaxed problem based on Equation~\eqref{eq:stoichiometric:equation:relaxed},
like the popular system \gapfill~\citep{SatishKumar2007}. Applied to the network of Fig.~\ref{gra:toy}, the minimal completion under the relaxed stoichiometric activation is $R''_1=\{r_{6}\}$  (Fig.~\ref{gra:toy_sr}) but does not carry flux because of the accumulation of metabolite $G$, allowed by Equation~\eqref{eq:stoichiometric:equation:relaxed}.
Note however that for strict steady-state modeling an \textit{a posteriori} verification of solutions is needed
to warrant the exact mass-balance equation~\eqref{eq:stoichiometric:equation}.

\begin{figure}
    \captionsetup{width=0.45\textwidth}
    \centering
    \begin{minipage}[t]{.5\textwidth}
      \centering
\usetikzlibrary{shapes.misc, positioning}
\begin{tikzpicture}[scale=0.45]\tiny
  \tikzstyle{metabolite}=[draw,circle,fill=white!80!black, text width=0.4cm, inner sep=0pt, align=center];
  \tikzstyle{repairmetabolite}=[draw,white!40!black, circle,fill=white!90!black,text=white!40!black,dashed];
  \tikzstyle{seed}=[draw,circle,fill=BlueGreen!70, text width=0.4cm, inner sep=0pt, align=center];
  \tikzstyle{target}=[draw,circle,fill=YellowOrange];
  \tikzstyle{reaction}=[draw,rectangle];
   \tikzstyle{export}=[draw,rectangle,dotted, very thick];
   \tikzstyle{exportrepair}=[draw,rectangle,dotted, very thick,white!80!black,text=white!70!black];
  \tikzstyle{repairreaction}=[draw,rectangle,white!40!black,text=white!40!black,dashed];
  \tikzstyle{solreaction}=[draw,rectangle,LimeGreen,text=black];
  \tikzstyle{initial}=[->,>=latex,thick];
  \tikzstyle{bdd}=[->,>=latex,thick];
  \tikzstyle{etiq}=[midway,fill=black!20,scale=0.5];
  \tikzstyle{stc}=[draw, rectangle, white, text=black]

  \node[stc] (stcr4C) at (6.2,6.7) {$2$};

  \draw [black,dotted, rounded corners, very thick] (-1,4.25) rectangle (12,9.5);

  \node[seed] (S2) at (0,7) {$S_{3}$};
  \node[metabolite] (Sb2) at (8.5,5.75) {$S_{2}$};
  \node[metabolite] (S1) at (0,8) {$S_{1}$};

  \node[metabolite] (F) at (11,7.50) {$F$};

  \node[metabolite] (A) at (6,8) {$A$};
  \node[metabolite] (B) at (2,5.0) {$B$};
  \node[metabolite] (C) at (7,6.5) {$C$};
  \node[metabolite] (D) at (5.50,4.75) {$D$};
  \node[metabolite] (E) at (4,6.5) {$E$};
  \node[repairmetabolite] (X) at (5,9) {$G$};

  \node[reaction] (R0) at (1.2,6.2) {$r_{0}$};
  \node[reaction] (R1) at (0.8,5.60) {$r_{1}$};
  \node[reaction] (R2) at (6.3,5.5) {$r_{2}$};
  \node[reaction] (R3) at (4.5,5.5) {$r_{3}$};
  \node[reaction] (R4) at (5.5,6.5) {$r_{4}$};
  \node[reaction, very thick,YellowOrange] (R5) at (9,7.50) {$r_{5}$}; 
  \node[solreaction] (R6) at (3,8) {$r_{6}$};
  \node[solreaction] (R9) at (8.5,8.5) {$r_{9}$};

  \draw[->,>=latex] (B.north west) -- (R0.south east);
  \draw[->,>=latex] (R0.north west) -- (S2.south east);

  \draw[->,>=latex] (S2.south east) -- (R1.north west);
  \draw[->,>=latex] (R1.south east) -- (B.west);

  \draw[->,>=latex] (Sb2.west) -- (7.5,5.70) -- (R2.east);
  \draw[] (C.south east) -- (7.5,5.70);
  \draw[->,>=latex] (R2.south) -- (D.east);

  \draw[->,>=latex] (D.west) -- (R3.south east);
  \draw[->,>=latex] (R3.north) -- (E.south east);

  \draw[->,>=latex] (E.east) -- (R4.west);
  \draw[->,>=latex] (R4.east) -- (C.west);

  \draw[->,>=latex] (A.east) -- (7.5,7.50) -- (R5.west);
  \draw[] (C.north east) -- (7.5,7.50);
  \draw[->,>=latex] (R5.east) -- (F.west);

  \draw[->,>=latex,LimeGreen] (S1.east) -- (R6.west);
  \draw[->,>=latex,LimeGreen] (R6.east) -- (4,8) -- (A.west);
  \draw[->,>=latex,LimeGreen] (4,8) -- (X.west);

  %

  \draw[->,>=latex,LimeGreen] (X.east) -- (R9.north west);
  \draw[->,>=latex,LimeGreen] (R9.east) -- (F.north west);


  \node[export] (outF) at (11,3.6) {\ExportReaction};
   \draw[->,>=stealth,white!30!black,dotted, very thick] (F.south) --  (outF.north);

   \node[export] (inS) at (0,10.2) {$r_{s_1}$};
   \draw[->,>=stealth,white!30!black,dotted, very thick] (inS.south) --  (S1.north);

  \node[export] (inS2) at (8.5,3.6) {$r_{s_2}$};
  \draw[->,>=stealth,white!30!black,dotted, very thick] (inS2) --  (Sb2.south);

\end{tikzpicture}
%
      \caption{Solution to metabolic network completion under stoichiometric activation hypothesis in order to satisfy Equations~\eqref{eq:stoichiometric:bounds},~\eqref{eq:stoichiometric:equation} and ~\eqref{eq:stoichiometric:activation}. Within this network, there exists at least one flux distribution which activates $r_5$.}
      \label{gra:toy_ss}
    \end{minipage}%
    \begin{minipage}[t]{.5\textwidth}
      \centering
\usetikzlibrary{shapes.misc, positioning}
\begin{tikzpicture}[scale=0.45]\tiny
  \tikzstyle{metabolite}=[draw,circle,fill=white!80!black, text width=0.4cm, inner sep=0pt, align=center];
  \tikzstyle{repairmetabolite}=[draw,white!40!black, circle,fill=white!90!black,text=white!40!black,dashed];
  \tikzstyle{seed}=[draw,circle,fill=BlueGreen!70, text width=0.4cm, inner sep=0pt, align=center];
  \tikzstyle{target}=[draw,circle,fill=YellowOrange];
  \tikzstyle{reaction}=[draw,rectangle];
   \tikzstyle{export}=[draw,rectangle,dotted, very thick];
   \tikzstyle{exportrepair}=[draw,rectangle,dotted, very thick,white!80!black,text=white!70!black];
  \tikzstyle{repairreaction}=[draw,rectangle,white!40!black,text=white!40!black,dashed];
  \tikzstyle{solreaction}=[draw,rectangle,LimeGreen,text=black];
  \tikzstyle{initial}=[->,>=latex,thick];
  \tikzstyle{bdd}=[->,>=latex,thick];
  \tikzstyle{etiq}=[midway,fill=black!20,scale=0.5];
  \tikzstyle{stc}=[draw, rectangle, white, text=black]

  \node[stc] (stcr4C) at (6.2,6.7) {$2$};

  \draw [black,dotted, rounded corners, very thick] (-1,4.25) rectangle (12,9.5);

  \node[seed] (S2) at (0,7) {$S_{3}$};
  \node[metabolite] (Sb2) at (8.5,5.75) {$S_{2}$};
  \node[metabolite] (S1) at (0,8) {$S_{1}$};

  \node[metabolite] (F) at (11,7.50) {$F$};

  \node[metabolite] (A) at (6,8) {$A$};
  \node[metabolite] (B) at (2,5.0) {$B$};
  \node[metabolite] (C) at (7,6.5) {$C$};
  \node[metabolite] (D) at (5.50,4.75) {$D$};
  \node[metabolite] (E) at (4,6.5) {$E$};
  \node[repairmetabolite] (X) at (5,9) {$G$};

  \node[reaction] (R0) at (1.2,6.2) {$r_{0}$};
  \node[reaction] (R1) at (0.8,5.60) {$r_{1}$};
  \node[reaction] (R2) at (6.3,5.5) {$r_{2}$};
  \node[reaction] (R3) at (4.5,5.5) {$r_{3}$};
  \node[reaction] (R4) at (5.5,6.5) {$r_{4}$};
  \node[reaction, very thick,YellowOrange] (R5) at (9,7.50) {$r_{5}$}; 
  \node[solreaction] (R6) at (3,8) {$r_{6}$};

  \draw[->,>=latex] (B.north west) -- (R0.south east);
  \draw[->,>=latex] (R0.north west) -- (S2.south east);

  \draw[->,>=latex] (S2.south east) -- (R1.north west);
  \draw[->,>=latex] (R1.south east) -- (B.west);

  \draw[->,>=latex] (Sb2.west) -- (7.5,5.70) -- (R2.east);
  \draw[] (C.south east) -- (7.5,5.70);
  \draw[->,>=latex] (R2.south) -- (D.east);

  \draw[->,>=latex] (D.west) -- (R3.south east);
  \draw[->,>=latex] (R3.north) -- (E.south east);

  \draw[->,>=latex] (E.east) -- (R4.west);
  \draw[->,>=latex] (R4.east) -- (C.west);

  \draw[->,>=latex] (A.east) -- (7.5,7.50) -- (R5.west);
  \draw[] (C.north east) -- (7.5,7.50);
  \draw[->,>=latex] (R5.east) -- (F.west);

  \draw[->,>=latex,LimeGreen] (S1.east) -- (R6.west);
  \draw[->,>=latex,LimeGreen] (R6.east) -- (4,8) -- (A.west);
  \draw[->,>=latex,LimeGreen] (4,8) -- (X.west);

  %



  \node[export] (outF) at (11,3.6) {\ExportReaction};
   \draw[->,>=stealth,white!30!black,dotted, very thick] (F.south) --  (outF.north);

   \node[export] (inS) at (0,10.2) {$r_{s_1}$};
   \draw[->,>=stealth,white!30!black,dotted, very thick] (inS.south) --  (S1.north);

  \node[export] (inS2) at (8.5,3.6) {$r_{s_2}$};
  \draw[->,>=stealth,white!30!black,dotted, very thick] (inS2) --  (Sb2.south);

\end{tikzpicture}
%
      \caption{Solution to metabolic network completion under relaxed stoichiometric activation hypothesis in order to satisfy Equations~\eqref{eq:stoichiometric:bounds},~\eqref{eq:stoichiometric:equation:relaxed} and ~\eqref{eq:stoichiometric:activation:relaxed}. Notice that within this completed network, there exist no flux distribution allowing the reaction $r_5$ to be activated.}
      \label{gra:toy_sr}
    \end{minipage}
\end{figure}

\subsection{Topological Metabolic Network Completion}\label{sec:topo} 
A qualitative approach to metabolic network completion relies on the topology of networks for capturing the activation of reactions.
Given a metabolic network $G$, a reaction $r\in R$ is \emph{activated} from a set of seeds $S$ if all reactants in $\Reactants{r}$ are reachable from~$S$.
Moreover, a metabolite $m\in M$ is \emph{reachable} from $S$ if %
$m\in S$
or if
$m\in\Products{r}$ for some reaction $r\in R$ where all $m'\in\Reactants{r}$ are reachable from~$S$.
The \emph{scope} of $S$, written $\Sigma_G(S)$, is the closure of compounds reachable from~$S$.
In this setting, \emph{topological activation} of reactions from a set of seeds $S$ is defined as follows:
\begin{align}
  r_T \in \Activity{t}{G}{S} \ \text{ iff } \ \Reactants{r_T} \subseteq \Sigma_G(S).   \label{eq:topological:activation}
\end{align} 
Note that this semantics avoids self-activated cycles by imposing an external entry sufficient to initiate all cycles ($S_3$ is not enough to activate the cycle as it does not activate one of its reaction on its own).
The resulting network completion problem can be expressed as a combinatorial optimization problem and effectively solved with ASP~\citep{schthi09a}.

For illustration, consider again the draft and reference networks $G$ and $G'$ in Fig.~\ref{gra:toy_d} and Fig.~\ref{gra:toy}.
We get $\Sigma_{G}(\{S_1,S_2,S_3\})=\{S_1,S_2,S_3,B\}$, indicating that target reaction $r_5$ is not activated from the seeds with the draft network
because $A$ and $C$, its reactants, are not reachable.
This changes once the network is completed.
Valid minimal completions are $R''_2=\{r_6,r_7\}$ (Fig.~\ref{gra:toy_st1}) and $R''_3=\{r_6,r_8\}$ (Fig.~\ref{gra:toy_st2}) because
\(
r_5\in\Activity{t}{G''_i}{\{S_1,S_2\}}\mbox{ since }\{A,C\}\subseteq\Sigma_{G''_i}(\{S_1,S_2\})
\)
for all extended networks $G''_i$ obtained from completions $R''_i$ of $G$ for $i\in\{2,3\}$.
%
\begin{figure}
    \captionsetup{width=0.45\textwidth}
    \centering
    \begin{minipage}[t]{.5\textwidth}
      \centering
\usetikzlibrary{shapes.misc, positioning}
\begin{tikzpicture}[scale=0.45]\tiny
  \tikzstyle{metabolite}=[draw,circle,fill=white!80!black, text width=0.4cm, inner sep=0pt, align=center];
  \tikzstyle{repairmetabolite}=[draw,white!40!black, circle,fill=white!90!black,text=white!40!black,dashed];
  \tikzstyle{seed}=[draw,circle,fill=BlueGreen!70, text width=0.4cm, inner sep=0pt, align=center];
  \tikzstyle{target}=[draw,circle,fill=YellowOrange];
  \tikzstyle{reaction}=[draw,rectangle];
   \tikzstyle{export}=[draw,rectangle,dotted, very thick];
   \tikzstyle{exportrepair}=[draw,rectangle,dotted, very thick,white!80!black,text=white!70!black];
  \tikzstyle{repairreaction}=[draw,rectangle,white!40!black,text=white!40!black,dashed];
  \tikzstyle{solreaction}=[draw,rectangle,LimeGreen,text=black];
  \tikzstyle{initial}=[->,>=latex,thick];
  \tikzstyle{bdd}=[->,>=latex,thick];
  \tikzstyle{etiq}=[midway,fill=black!20,scale=0.5];
  \tikzstyle{stc}=[draw, rectangle, white, text=black]

  \node[stc] (stcr4C) at (6.2,6.7) {$2$};

  \draw [black,dotted, rounded corners, very thick] (-1,4.25) rectangle (12,9.5);

  \node[seed] (S2) at (0,7) {$S_{3}$};
  \node[metabolite] (Sb2) at (8.5,5.75) {$S_{2}$};
  \node[metabolite] (S1) at (0,8) {$S_{1}$};

  \node[metabolite] (F) at (11,7.50) {$F$};

  \node[metabolite] (A) at (6,8) {$A$};
  \node[metabolite] (B) at (2,5.0) {$B$};
  \node[metabolite] (C) at (7,6.5) {$C$};
  \node[metabolite] (D) at (5.50,4.75) {$D$};
  \node[metabolite] (E) at (4,6.5) {$E$};
  \node[repairmetabolite] (X) at (5,9) {$G$};

  \node[reaction] (R0) at (1.2,6.2) {$r_{0}$};
  \node[reaction] (R1) at (0.8,5.60) {$r_{1}$};
  \node[reaction] (R2) at (6.3,5.5) {$r_{2}$};
  \node[reaction] (R3) at (4.5,5.5) {$r_{3}$};
  \node[reaction] (R4) at (5.5,6.5) {$r_{4}$};
  \node[reaction, very thick,YellowOrange] (R5) at (9,7.50) {$r_{5}$}; 
  \node[solreaction] (R6) at (3,8) {$r_{6}$};
  \node[solreaction] (R7) at (2.3,6.9) {$r_{7}$};

  \draw[->,>=latex] (B.north west) -- (R0.south east);
  \draw[->,>=latex] (R0.north west) -- (S2.south east);

  \draw[->,>=latex] (S2.south east) -- (R1.north west);
  \draw[->,>=latex] (R1.south east) -- (B.west);

  \draw[->,>=latex] (Sb2.west) -- (7.5,5.70) -- (R2.east);
  \draw[] (C.south east) -- (7.5,5.70);
  \draw[->,>=latex] (R2.south) -- (D.east);

  \draw[->,>=latex] (D.west) -- (R3.south east);
  \draw[->,>=latex] (R3.north) -- (E.south east);

  \draw[->,>=latex] (E.east) -- (R4.west);
  \draw[->,>=latex] (R4.east) -- (C.west);

  \draw[->,>=latex] (A.east) -- (7.5,7.50) -- (R5.west);
  \draw[] (C.north east) -- (7.5,7.50);
  \draw[->,>=latex] (R5.east) -- (F.west);

  \draw[->,>=latex,LimeGreen] (S1.east) -- (R6.west);
  \draw[->,>=latex,LimeGreen] (R6.east) -- (4,8) -- (A.west);
  \draw[->,>=latex,LimeGreen] (4,8) -- (X.west);

  \draw[->,>=latex,LimeGreen] (S2.east) -- (R7.west);
  \draw[->,>=latex,LimeGreen] (R7.east) -- (E.north west);
  %



  \node[export] (outF) at (11,3.6) {\ExportReaction};
   \draw[->,>=stealth,white!30!black,dotted, very thick] (F.south) --  (outF.north);

   \node[export] (inS) at (0,10.2) {$r_{s_1}$};
   \draw[->,>=stealth,white!30!black,dotted, very thick] (inS.south) --  (S1.north);

  \node[export] (inS2) at (8.5,3.6) {$r_{s_2}$};
  \draw[->,>=stealth,white!30!black,dotted, very thick] (inS2) --  (Sb2.south);

\end{tikzpicture}
%
      \caption{First solution to metabolic network completion under topological activation hypothesis satisfying Equation~\eqref{eq:topological:activation}. The production of C cannot be explained by a self-activated cycle and requires an external source of compounds via $S_3$ and reaction $r_7$.}
      \label{gra:toy_st1}
    \end{minipage}%
    \begin{minipage}[t]{.5\textwidth}
      \centering
\usetikzlibrary{shapes.misc, positioning}
\begin{tikzpicture}[scale=0.45]\tiny
  \tikzstyle{metabolite}=[draw,circle,fill=white!80!black, text width=0.4cm, inner sep=0pt, align=center];
  \tikzstyle{repairmetabolite}=[draw,white!40!black, circle,fill=white!90!black,text=white!40!black,dashed];
  \tikzstyle{seed}=[draw,circle,fill=BlueGreen!70, text width=0.4cm, inner sep=0pt, align=center];
  \tikzstyle{target}=[draw,circle,fill=YellowOrange];
  \tikzstyle{reaction}=[draw,rectangle];
   \tikzstyle{export}=[draw,rectangle,dotted, very thick];
   \tikzstyle{exportrepair}=[draw,rectangle,dotted, very thick,white!80!black,text=white!70!black];
  \tikzstyle{repairreaction}=[draw,rectangle,white!40!black,text=white!40!black,dashed];
  \tikzstyle{solreaction}=[draw,rectangle,LimeGreen,text=black];
  \tikzstyle{initial}=[->,>=latex,thick];
  \tikzstyle{bdd}=[->,>=latex,thick];
  \tikzstyle{etiq}=[midway,fill=black!20,scale=0.5];
  \tikzstyle{stc}=[draw, rectangle, white, text=black]

  \node[stc] (stcr4C) at (6.2,6.7) {$2$};

  \draw [black,dotted, rounded corners, very thick] (-1,4.25) rectangle (12,9.5);

  \node[seed] (S2) at (0,7) {$S_{3}$};
  \node[metabolite] (Sb2) at (8.5,5.75) {$S_{2}$};
  \node[metabolite] (S1) at (0,8) {$S_{1}$};

  \node[metabolite] (F) at (11,7.50) {$F$};

  \node[metabolite] (A) at (6,8) {$A$};
  \node[metabolite] (B) at (2,5.0) {$B$};
  \node[metabolite] (C) at (7,6.5) {$C$};
  \node[metabolite] (D) at (5.50,4.75) {$D$};
  \node[metabolite] (E) at (4,6.5) {$E$};
  \node[repairmetabolite] (X) at (5,9) {$G$};

  \node[reaction] (R0) at (1.2,6.2) {$r_{0}$};
  \node[reaction] (R1) at (0.8,5.60) {$r_{1}$};
  \node[reaction] (R2) at (6.3,5.5) {$r_{2}$};
  \node[reaction] (R3) at (4.5,5.5) {$r_{3}$};
  \node[reaction] (R4) at (5.5,6.5) {$r_{4}$};
  \node[reaction, very thick,YellowOrange] (R5) at (9,7.50) {$r_{5}$}; 
  \node[solreaction] (R6) at (3,8) {$r_{6}$};
  \node[solreaction] (R8) at (3,5.8) {$r_{8}$};

  \draw[->,>=latex] (B.north west) -- (R0.south east);
  \draw[->,>=latex] (R0.north west) -- (S2.south east);

  \draw[->,>=latex] (S2.south east) -- (R1.north west);
  \draw[->,>=latex] (R1.south east) -- (B.west);

  \draw[->,>=latex] (Sb2.west) -- (7.5,5.70) -- (R2.east);
  \draw[] (C.south east) -- (7.5,5.70);
  \draw[->,>=latex] (R2.south) -- (D.east);

  \draw[->,>=latex] (D.west) -- (R3.south east);
  \draw[->,>=latex] (R3.north) -- (E.south east);

  \draw[->,>=latex] (E.east) -- (R4.west);
  \draw[->,>=latex] (R4.east) -- (C.west);

  \draw[->,>=latex] (A.east) -- (7.5,7.50) -- (R5.west);
  \draw[] (C.north east) -- (7.5,7.50);
  \draw[->,>=latex] (R5.east) -- (F.west);

  \draw[->,>=latex,LimeGreen] (S1.east) -- (R6.west);
  \draw[->,>=latex,LimeGreen] (R6.east) -- (4,8) -- (A.west);
  \draw[->,>=latex,LimeGreen] (4,8) -- (X.west);


  \draw[->,>=latex,LimeGreen] (B.north east) -- (R8.south west);
  \draw[->,>=latex,LimeGreen] (R8.east) -- (E.south);



  \node[export] (outF) at (11,3.6) {\ExportReaction};
   \draw[->,>=stealth,white!30!black,dotted, very thick] (F.south) --  (outF.north);

   \node[export] (inS) at (0,10.2) {$r_{s_1}$};
   \draw[->,>=stealth,white!30!black,dotted, very thick] (inS.south) --  (S1.north);

  \node[export] (inS2) at (8.5,3.6) {$r_{s_2}$};
  \draw[->,>=stealth,white!30!black,dotted, very thick] (inS2) --  (Sb2.south);

\end{tikzpicture}
%
      \caption{Second solution to metabolic network completion under topological activation hypothesis satisfying Equation~\eqref{eq:topological:activation}.}
      \label{gra:toy_st2}
    \end{minipage}
\end{figure}

Relevant elements from the reference network are given in dashed gray.

\subsection{Hybrid Metabolic Network Completion}\label{sec:hybrid} 
The idea of hybrid metabolic network completion is to combine the two previous activation semantics:
the topological one accounts for a well-founded initiation of the system from the seeds
and the stoichiometric one warrants its mass-balance.
We thus aim at network completions that are both topologically functional and flux balanced
(without suffering from self-activated cycles).
More precisely,
a reaction $r_T\in R_T$ is \emph{hybridly activated} from a set $S$ of seeds in a network $G$,
if both criteria apply:
\begin{align}
\label{eq:hybrid:activation}
r_T \in \Activity{h}{G}{S} \ \text{ iff } \ r_T \in \Activity{s}{G}{S}\text{ and }r_T \in \Activity{t}{G}{S}.
\end{align}

Applying this to our example in Fig.~\ref{gra:toy},
we get the (minimal) hybrid solutions $R''_4=\{r_6,r_7,r_{9}\}$ (Fig.~\ref{gra:toy_sh1}) and $R''_5=\{r_6,r_8,r_{9}\}$ (Fig.~\ref{gra:toy_sh2}).
Both (topologically) initiate paths of reactions from the seeds to the target,
ie.\ $r_5\in\Activity{t}{G''_i}{\{S_1,S_2,S_3\}}\mbox{ since }\{A,C\}\subseteq\Sigma_{G''_i}(\{S_1,S_2,S_3\})$
for both extended networks $G''_i$ obtained from completions $R''_i$ of $G$ for $i\in\{4,5\}$.
Both solutions are as well stoichiometrically valid and balance the amount of every metabolite,
hence we also have $r_5\in\Activity{s}{G''_i}{\{S_1,S_2,S_3\}}$.

\begin{figure}
    \captionsetup{width=0.45\textwidth}
    \centering
    \begin{minipage}[t]{.50\textwidth}
      \centering
\usetikzlibrary{shapes.misc, positioning}
\begin{tikzpicture}[scale=0.45]\tiny
  \tikzstyle{metabolite}=[draw,circle,fill=white!80!black, text width=0.4cm, inner sep=0pt, align=center];
  \tikzstyle{repairmetabolite}=[draw,white!40!black, circle,fill=white!90!black,text=white!40!black,dashed];
  \tikzstyle{seed}=[draw,circle,fill=BlueGreen!70, text width=0.4cm, inner sep=0pt, align=center];
  \tikzstyle{target}=[draw,circle,fill=YellowOrange];
  \tikzstyle{reaction}=[draw,rectangle];
   \tikzstyle{export}=[draw,rectangle,dotted, very thick];
   \tikzstyle{exportrepair}=[draw,rectangle,dotted, very thick,white!80!black,text=white!70!black];
  \tikzstyle{repairreaction}=[draw,rectangle,white!40!black,text=white!40!black,dashed];
  \tikzstyle{solreaction}=[draw,rectangle,LimeGreen,text=black];
  \tikzstyle{initial}=[->,>=latex,thick];
  \tikzstyle{bdd}=[->,>=latex,thick];
  \tikzstyle{etiq}=[midway,fill=black!20,scale=0.5];
  \tikzstyle{stc}=[draw, rectangle, white, text=black]

  \node[stc] (stcr4C) at (6.2,6.7) {$2$};

  \draw [black,dotted, rounded corners, very thick] (-1,4.25) rectangle (12,9.5);

  \node[seed] (S2) at (0,7) {$S_{3}$};
  \node[metabolite] (Sb2) at (8.5,5.75) {$S_{2}$};
  \node[metabolite] (S1) at (0,8) {$S_{1}$};

  \node[metabolite] (F) at (11,7.50) {$F$};

  \node[metabolite] (A) at (6,8) {$A$};
  \node[metabolite] (B) at (2,5.0) {$B$};
  \node[metabolite] (C) at (7,6.5) {$C$};
  \node[metabolite] (D) at (5.50,4.75) {$D$};
  \node[metabolite] (E) at (4,6.5) {$E$};
  \node[repairmetabolite] (X) at (5,9) {$G$};

  \node[reaction] (R0) at (1.2,6.2) {$r_{0}$};
  \node[reaction] (R1) at (0.8,5.60) {$r_{1}$};
  \node[reaction] (R2) at (6.3,5.5) {$r_{2}$};
  \node[reaction] (R3) at (4.5,5.5) {$r_{3}$};
  \node[reaction] (R4) at (5.5,6.5) {$r_{4}$};
  \node[reaction, very thick,YellowOrange] (R5) at (9,7.50) {$r_{5}$}; 
  \node[solreaction] (R6) at (3,8) {$r_{6}$};
  \node[solreaction] (R7) at (2.3,6.9) {$r_{7}$};
  \node[solreaction] (R9) at (8.5,8.5) {$r_{9}$};

  \draw[->,>=latex] (B.north west) -- (R0.south east);
  \draw[->,>=latex] (R0.north west) -- (S2.south east);

  \draw[->,>=latex] (S2.south east) -- (R1.north west);
  \draw[->,>=latex] (R1.south east) -- (B.west);

  \draw[->,>=latex] (Sb2.west) -- (7.5,5.70) -- (R2.east);
  \draw[] (C.south east) -- (7.5,5.70);
  \draw[->,>=latex] (R2.south) -- (D.east);

  \draw[->,>=latex] (D.west) -- (R3.south east);
  \draw[->,>=latex] (R3.north) -- (E.south east);

  \draw[->,>=latex] (E.east) -- (R4.west);
  \draw[->,>=latex] (R4.east) -- (C.west);

  \draw[->,>=latex] (A.east) -- (7.5,7.50) -- (R5.west);
  \draw[] (C.north east) -- (7.5,7.50);
  \draw[->,>=latex] (R5.east) -- (F.west);

  \draw[->,>=latex,LimeGreen] (S1.east) -- (R6.west);
  \draw[->,>=latex,LimeGreen] (R6.east) -- (4,8) -- (A.west);
  \draw[->,>=latex,LimeGreen] (4,8) -- (X.west);

  \draw[->,>=latex,LimeGreen] (S2.east) -- (R7.west);
  \draw[->,>=latex,LimeGreen] (R7.east) -- (E.north west);
  %

  \draw[->,>=latex,LimeGreen] (X.east) -- (R9.north west);
  \draw[->,>=latex,LimeGreen] (R9.east) -- (F.north west);


  \node[export] (outF) at (11,3.6) {\ExportReaction};
   \draw[->,>=stealth,white!30!black,dotted, very thick] (F.south) --  (outF.north);

   \node[export] (inS) at (0,10.2) {$r_{s_1}$};
   \draw[->,>=stealth,white!30!black,dotted, very thick] (inS.south) --  (S1.north);

  \node[export] (inS2) at (8.5,3.6) {$r_{s_2}$};
  \draw[->,>=stealth,white!30!black,dotted, very thick] (inS2) --  (Sb2.south);

\end{tikzpicture}
%
      \caption{First solution to metabolic network completion under hybrid activation hypothesis satisfying Equation~\eqref{eq:hybrid:activation} (that is Equations~\eqref{eq:stoichiometric:bounds},~\eqref{eq:stoichiometric:equation}, ~\eqref{eq:stoichiometric:activation} and ~\eqref{eq:topological:activation}).}
      \label{gra:toy_sh1}
    \end{minipage}%
    \begin{minipage}[t]{.50\textwidth}
      \centering
\usetikzlibrary{shapes.misc, positioning}
\begin{tikzpicture}[scale=0.45]\tiny
  \tikzstyle{metabolite}=[draw,circle,fill=white!80!black, text width=0.4cm, inner sep=0pt, align=center];
  \tikzstyle{repairmetabolite}=[draw,white!40!black, circle,fill=white!90!black,text=white!40!black,dashed];
  \tikzstyle{seed}=[draw,circle,fill=BlueGreen!70, text width=0.4cm, inner sep=0pt, align=center];
  \tikzstyle{target}=[draw,circle,fill=YellowOrange];
  \tikzstyle{reaction}=[draw,rectangle];
   \tikzstyle{export}=[draw,rectangle,dotted, very thick];
   \tikzstyle{exportrepair}=[draw,rectangle,dotted, very thick,white!80!black,text=white!70!black];
  \tikzstyle{repairreaction}=[draw,rectangle,white!40!black,text=white!40!black,dashed];
  \tikzstyle{solreaction}=[draw,rectangle,LimeGreen,text=black];
  \tikzstyle{initial}=[->,>=latex,thick];
  \tikzstyle{bdd}=[->,>=latex,thick];
  \tikzstyle{etiq}=[midway,fill=black!20,scale=0.5];
  \tikzstyle{stc}=[draw, rectangle, white, text=black]

  \node[stc] (stcr4C) at (6.2,6.7) {$2$};

  \draw [black,dotted, rounded corners, very thick] (-1,4.25) rectangle (12,9.5);

  \node[seed] (S2) at (0,7) {$S_{3}$};
  \node[metabolite] (Sb2) at (8.5,5.75) {$S_{2}$};
  \node[metabolite] (S1) at (0,8) {$S_{1}$};

  \node[metabolite] (F) at (11,7.50) {$F$};

  \node[metabolite] (A) at (6,8) {$A$};
  \node[metabolite] (B) at (2,5.0) {$B$};
  \node[metabolite] (C) at (7,6.5) {$C$};
  \node[metabolite] (D) at (5.50,4.75) {$D$};
  \node[metabolite] (E) at (4,6.5) {$E$};
  \node[repairmetabolite] (X) at (5,9) {$G$};

  \node[reaction] (R0) at (1.2,6.2) {$r_{0}$};
  \node[reaction] (R1) at (0.8,5.60) {$r_{1}$};
  \node[reaction] (R2) at (6.3,5.5) {$r_{2}$};
  \node[reaction] (R3) at (4.5,5.5) {$r_{3}$};
  \node[reaction] (R4) at (5.5,6.5) {$r_{4}$};
  \node[reaction, very thick,YellowOrange] (R5) at (9,7.50) {$r_{5}$}; 
  \node[solreaction] (R6) at (3,8) {$r_{6}$};
  \node[solreaction] (R8) at (3,5.8) {$r_{8}$};
  \node[solreaction] (R9) at (8.5,8.5) {$r_{9}$};

  \draw[->,>=latex] (B.north west) -- (R0.south east);
  \draw[->,>=latex] (R0.north west) -- (S2.south east);

  \draw[->,>=latex] (S2.south east) -- (R1.north west);
  \draw[->,>=latex] (R1.south east) -- (B.west);

  \draw[->,>=latex] (Sb2.west) -- (7.5,5.70) -- (R2.east);
  \draw[] (C.south east) -- (7.5,5.70);
  \draw[->,>=latex] (R2.south) -- (D.east);

  \draw[->,>=latex] (D.west) -- (R3.south east);
  \draw[->,>=latex] (R3.north) -- (E.south east);

  \draw[->,>=latex] (E.east) -- (R4.west);
  \draw[->,>=latex] (R4.east) -- (C.west);

  \draw[->,>=latex] (A.east) -- (7.5,7.50) -- (R5.west);
  \draw[] (C.north east) -- (7.5,7.50);
  \draw[->,>=latex] (R5.east) -- (F.west);

  \draw[->,>=latex,LimeGreen] (S1.east) -- (R6.west);
  \draw[->,>=latex,LimeGreen] (R6.east) -- (4,8) -- (A.west);
  \draw[->,>=latex,LimeGreen] (4,8) -- (X.west);


  \draw[->,>=latex,LimeGreen] (B.north east) -- (R8.south west);
  \draw[->,>=latex,LimeGreen] (R8.east) -- (E.south);

  \draw[->,>=latex,LimeGreen] (X.east) -- (R9.north west);
  \draw[->,>=latex,LimeGreen] (R9.east) -- (F.north west);


  \node[export] (outF) at (11,3.6) {\ExportReaction};
   \draw[->,>=stealth,white!30!black,dotted, very thick] (F.south) --  (outF.north);

   \node[export] (inS) at (0,10.2) {$r_{s_1}$};
   \draw[->,>=stealth,white!30!black,dotted, very thick] (inS.south) --  (S1.north);

  \node[export] (inS2) at (8.5,3.6) {$r_{s_2}$};
  \draw[->,>=stealth,white!30!black,dotted, very thick] (inS2) --  (Sb2.south);

\end{tikzpicture}
%
      \caption{Second solution to metabolic network completion under hybrid activation hypothesis satisfying Equation~\eqref{eq:hybrid:activation} (that is Equations~\eqref{eq:stoichiometric:bounds},~\eqref{eq:stoichiometric:equation}, ~\eqref{eq:stoichiometric:activation} and ~\eqref{eq:topological:activation}).}
      \label{gra:toy_sh2}
    \end{minipage}
\end{figure}

\subsection{Union of Metabolic Network Completions}\label{sec:union} 
As depicted in the toy examples for the topological (Fig.~\ref{gra:toy_st1} and Fig.~\ref{gra:toy_st2}) and hybrid (Fig.~\ref{gra:toy_sh1} and
Fig.~\ref{gra:toy_sh2}) activation, several minimal solutions to one metabolic network completion problem may exist.
There might be dozens of minimal completions, depending on the degradation of the original draft network,
hence leading to difficulties for biologists and bioinformaticians to discriminate the individual results.
One solution to facilitate this curation task is to provide, in addition to the enumeration of solutions, their union.
This has been done previously for the topological completion \citep{Prigent2017}.

Notably, the concept of ``union of solutions" is particularly relevant from the biological perspective since it provides in a single view all possible reactions that could be inserted in a solution to the network completion problem.
Additionally, verifying the union according to the desired (stoichiometric and hybrid) activation semantics,
offers a way to analyze the quality of approximation methods (topological and relaxed-stoichiometric ones).
If individual solutions contradict a definition of activation that the union satisfies, it suggests that the family of reactions contained in the union, although possibly non-minimal, may be of interest.
Thus providing merit to the approximation method and their results.

Importantly, we notice that the operation of performing the union of solutions is stable with the concept of activation, although it can contradict the minimality of the size of completion.
Indeed, the union of solutions to the topological network completion problem is itself a (non-minimal) solution to the topological completion problem.
Similarly, the union of minimal stoichiometric solutions always displays the stoichiometric activation of the target reaction(s).
In fact, adding an arbitrary set of reactions to a metabolic network still maintains stoichiometric activation,
since flux distribution for the newly added reactions may be set to zero.
Consequently, the union of minimal hybrid solutions always displays the hybrid activation in the target reaction(s).

The following theorems (Theorems ~\ref{th:topo}, ~\ref{th:flux} and ~\ref{th:hybr}) are a formalization of the stability of the union of solutions with respect to the three concepts of activation.

The union $G=G_1\cup G_2$ of two metabolic networks $G_1=(R_1\cup M_1,E_1,\StoichiometricFunction_1)$
and $G_2=(R_2\cup M_2,E_2,\StoichiometricFunction_2)$ is defined by
\begin{align}
G &= (R\cup M, E, \StoichiometricFunction), \label{eq:graph.union1}\\
R &= R_1\cup R_2, \label{eq:graph.union2}\\
M &= M_1\cup M_2, \label{eq:graph.union3}\\
E &= E_1\cup E_2, \label{eq:graph.union4}\\
\StoichiometricFunction &= \StoichiometricFunction_1\cup \StoichiometricFunction_2. \label{eq:graph.union5}
\end{align}

\begin{theorem}\label{th:topo}
Let $G_1$ and $G_2$ be metabolic networks.
If $R_T\subseteq\Activity{t}{G_1}{S}$, then $R_T\subseteq\Activity{t}{G_1\cup G_2}{S}$.
\end{theorem}

\begin{proof}
The proof is given by monotonicity of the union and the monotonicity of the closure.
Thus it can never be case that having more reactions disables reachability.
More formal,
$R_T\subseteq\Activity{t}{G_1}{S}$ holds iff
$\Reactants{r_T}\subseteq\Sigma_{G_1}(S)$.
Furthermore, we have
$\Sigma_{G_1}(S)\subseteq \Sigma_{G_1\cup G_2}(S)$
by the definition of the closure.
This implies
$\Reactants{r_T}\subseteq\Sigma_{G_1\cup G_2}(S)$.
Finally, we have
$R_T\subseteq\Activity{t}{G_1\cup G_2}{S}$.
\end{proof}

\begin{theorem}\label{th:flux}
Let $G_1$ and $G_2$ be metabolic networks.
If $R_T\subseteq\Activity{s}{G_1}{S}$, then $R_T\subseteq\Activity{s}{G_1\cup G_2}{S}$.
\end{theorem}

\begin{proof}
First, we define following bijective functions
\begin{align*}
f:&R_1 \rightarrow \{1,\dots,l\}\subseteq\mathbb{N}, \\
& r \mapsto f(r)=i \\
g:&M_1 \rightarrow \{1,\dots,k\}\subseteq\mathbb{N}, \\
& m \mapsto g(m)=j \\
f':&R_1\cup R_2 \rightarrow \{1,\dots,l'\}\subseteq\mathbb{N}, \\
& r \mapsto f'(r)=
    \begin{cases}
    f(r) &, \text{ if $f(r)$ is defined} \\
    i    &, \text{ otherwise}
    \end{cases} \\
g'&:M_1\cup M_2 \rightarrow \{1,\dots,k'\}\subseteq\mathbb{N} \\
& m \mapsto g'(m)=
    \begin{cases}
    g(m) &, \text{ if $g(m)$ is defined} \\
    j    &, \text{ otherwise}
    \end{cases}
\end{align*}
for $k=|M_1|$, $l=|R_1|$, $k'=|M_1\cup M_2|$ and $l'=|R_1\cup R_2|$ regarding $G_1$ and $G_1\cup G_2$, respectively.
Now, we rewrite the system of (\ref{eq:stoichiometric:equation}) regarding $G_1$ as a matrix equation $Av=0$ of form
\begin{align*}
\begin{pmatrix}
a_{11} & \dots & a_{1l} \\
\vdots & \ddots & \vdots \\
a_{k1} & \dots & a_{kl}
\end{pmatrix}
\begin{pmatrix}
v_1 \\
\vdots  \\
v_l
\end{pmatrix}
=
\begin{pmatrix}
0 \\
\vdots \\
0
\end{pmatrix}\label{eq:matrix}
\end{align*}
where $A$ is a $k\times l$ matrix with coefficients
\begin{align*}
a_{g(m)f(r)}=
    \begin{cases}
    \StoichiometricFunction_1(r,m)  &, (r,m)\in E_1 \\
    -\StoichiometricFunction_1(m,r) &, (m,r)\in E_1 \\
    0       &, \text{ otherwise}
    \end{cases}
\end{align*}
and $v$ consists of variables $v_{f(r)}$ for $r\in R_1$.
By $L=\{v\mid Av=0\}$ we denote the set of solutions induced by $Av=0$.

Furthermore, we represent the system of linear equations of (\ref{eq:stoichiometric:equation}) regarding $G_1\cup G_2$ as a matrix equation $A'v'=0$ of form
\begin{align*}
\begin{pmatrix}
a_{11} & \dots & a_{1l} & a_{1l+1} & \dots & a_{1l'} \\
\vdots & \ddots& \vdots & \vdots   & \ddots& \vdots \\
a_{k1} & \dots & a_{kl} & a_{kl+1} & \dots & a_{kl'} \\
0      & \dots & 0      & a_{k+1l+1} & \dots & a_{k+1l'} \\
\vdots & \ddots& \vdots & \vdots   & \ddots& \vdots \\
0      & \dots & 0      & a_{k'l+1}& \dots & a_{k'l'}
\end{pmatrix}
\begin{pmatrix}
v_1 \\
\vdots  \\
v_l \\
v_{l+1} \\
\vdots \\
v_{l'}
\end{pmatrix}
=
\begin{pmatrix}
0 \\
\vdots \\
0
\end{pmatrix}
\end{align*}
where $A'$ is a $k'\times l'$ matrix with coefficients
\begin{align*}
a_{g'(m)f'(r)}=
    \begin{cases}
    \StoichiometricFunction(r,m)  &, (r,m)\in E_1\cup E_2 \\
    -\StoichiometricFunction(m,r) &, (m,r)\in E_1\cup E_2 \\
    0       &, \text{ otherwise}
    \end{cases}
\end{align*}
where $\StoichiometricFunction=\StoichiometricFunction_1\cup \StoichiometricFunction_2$
and $v'$ consists of variables $v_{f'(r)}$ of (\ref{eq:stoichiometric:equation}) for $r\in R_1\cup R_2$.
Note that $A'$ can always be written in this form, since switching columns and rows will not change solutions.
By $L'=\{v'\mid A'v'=0\}$ we denote the set of solutions induced by $A'v'=0$.

Since $A'v'=0$ is homogeneous, $L\subseteq L'$ holds by extending $L$ with zeros for $v_{f'(r)}$ with $r\in R_2\setminus R_1$.
Thus $\{v\mid v\in L, \forall r_T\in R_T, v_{f(r_T)}>0\}
\subseteq\{v\mid v\in L', \forall r_T\in R_T, v_{f'(r_T)}>0\}$
by extending the first set with zeros for $v_{f'(r)}$ with $r\in R_2\setminus R_1$.
From $R_T\subseteq\Activity{s}{G_1}{S}$, we know that the homogeneous system of linear equations from
(\ref{eq:stoichiometric:equation}) regarding $G_1$ is non-trivial satisfiable,
which finally implies that $R_T\subseteq\Activity{s}{G_1\cup G_2}{S}$.
\end{proof}

\begin{theorem}\label{th:hybr}
Let $G_1$ and $G_2$ be metabolic networks.
If $R_T\subseteq\Activity{h}{G_1}{S}$, then $R_T\subseteq\Activity{h}{G_1\cup G_2}{S}$.
\end{theorem}

\begin{proof}
Follows directly by the definition of hybrid activation together with Theorem~\ref{th:topo} and Theorem~\ref{th:flux}.
More formal,
$R_T\subseteq\Activity{h}{G_1}{S}$
holds iff
$R_T\subseteq\Activity{t}{G_1}{S}$ and $R_T\subseteq\Activity{s}{G_1}{S}$.
From Theorem~\ref{th:topo} and $R_T\subseteq\Activity{t}{G_1}{S}$ follows
$R_T\subseteq\Activity{t}{G_1\cup G_2}{S}$.
Analogously, from Theorem~\ref{th:flux} and $R_T\subseteq\Activity{s}{G_1}{S}$ follows
$R_T\subseteq\Activity{s}{G_1\cup G_2}{S}$.
Finally, this implies
$R_T\subseteq\Activity{h}{G_1\cup G_2}{S}$.
\end{proof}

In particular, studying the union in case of topological modeling can pinpoint interesting cases.
Individual solutions satisfying the topological activation can additionally satisfy the stoichiometric and thus the hybrid activation semantics.
A union including such a solution will also adhere to the hybrid standard.
In some cases, the union of solutions will display the stoichiometric activation whereas the individual solutions only satisfy the topological activation.
Fig.~\ref{gra:union_nf_f1} to Fig.~\ref{gra:union_nf_f} display an example of topological metabolic network completions that do not satisfy stoichiometric (and hybrid) activation whereas their union does.
Fig.~\ref{gra:union_nf_nf1} to Fig.~\ref{gra:union_nf_nf} provide an example of minimal topological completions that do not satisfy stoichiometric (and hybrid) activation and for which the union does not satisfy it either.
%

Both observations induce that in general we cannot derive anything about activation of reactions in a graph resulting from the union of two or more graphs.
And similarly, we cannot infer about the activation of reactions in subgraphs arbitrarily derived from a graph in which these reactions are activated.

\begin{figure}
    \captionsetup{width=0.3\textwidth}
    \centering
    \begin{minipage}[t]{.32\textwidth}
%
%


    \usetikzlibrary{shapes.misc, positioning}
    \begin{tikzpicture}[scale=0.39]\tiny
      \tikzstyle{metabolite}=[draw,circle,fill=white!80!black];
      \tikzstyle{repairmetabolite}=[draw,white!40!black, circle,fill=white!90!black,text=white!40!black,dashed];
      \tikzstyle{seed}=[draw,circle,fill=BlueGreen!70];
      \tikzstyle{target}=[draw,circle,fill=YellowOrange];
      \tikzstyle{reaction}=[draw,rectangle];
       \tikzstyle{export}=[draw,rectangle,dotted, very thick];
       \tikzstyle{exportrepair}=[draw,rectangle,dotted, very thick,white!80!black,text=white!70!black];
      \tikzstyle{repairreaction}=[draw,rectangle,white!40!black,text=white!40!black,dashed];
      \tikzstyle{solreaction}=[draw,rectangle,LimeGreen,text=black];
      \tikzstyle{initial}=[->,>=latex,thick];
      \tikzstyle{bdd}=[->,>=latex,thick];
      \tikzstyle{etiq}=[midway,fill=black!20,scale=0.5];
      \tikzstyle{stc}=[draw, rectangle, white, text=black]

      \draw [black,dotted, rounded corners, very thick] (-0.5,4) rectangle (10.2,9);

      \node[seed] (S) at (0,7) {$S$};
      \node[metabolite] (A) at (3,8) {$A$};
      \node[metabolite] (B) at (3,6) {$B$};
      \node[metabolite] (C) at (6.2,7) {$C$};
      \node[metabolite] (D) at (9.5,7) {$D$};

      \node[reaction] (R1) at (1.5,7) {$r_{1}$};
      \node[reaction, very thick,YellowOrange] (R4) at (7.7,7) {$r_{4}$}; 
      \node[solreaction] (R2) at (4.7,7.6) {$r_{2}$};

      \draw[->,>=latex] (S.east) -- (R1.west);
      \draw[->,>=latex] (R1.east) -- (2.2,7) -- (A.west);
      \draw[->,>=latex] (2.2,7)  -- (B.west);

      \draw[->,>=latex,LimeGreen] (A.east) -- (R2.west);
      \draw[->,>=latex,LimeGreen] (R2.east) -- (C.north west);


      \draw[->,>=latex] (C.east) -- (R4.west);
      \draw[->,>=latex] (R4.east) -- (D.west);

      \node[export] (outD) at (9.5,3) {\ExportReaction};
      \draw[->,>=stealth,white!30!black,dotted, very thick] (D.south) --  (outD.north);

      \node[export] (inS) at (0,10) {$r_{s}$};
      \draw[->,>=stealth,white!30!black,dotted, very thick] (inS) --  (S.north);

  \end{tikzpicture}

      \caption{Topological completion $R_1=\{r_2\}$ satisfies $r_4\in\Activity{t}{G_1}{\{S\}}$, but carries no flux, due to accumulation of compound $B$ that contradicts Eq.~\ref{eq:stoichiometric:equation}.\label{gra:union_nf_f1}}
    \end{minipage}
    \begin{minipage}[t]{.32\textwidth}
%
%


    \usetikzlibrary{shapes.misc, positioning}
    \begin{tikzpicture}[scale=0.39]\tiny
      \tikzstyle{metabolite}=[draw,circle,fill=white!80!black];
      \tikzstyle{repairmetabolite}=[draw,white!40!black, circle,fill=white!90!black,text=white!40!black,dashed];
      \tikzstyle{seed}=[draw,circle,fill=BlueGreen!70];
      \tikzstyle{target}=[draw,circle,fill=YellowOrange];
      \tikzstyle{reaction}=[draw,rectangle];
       \tikzstyle{export}=[draw,rectangle,dotted, very thick];
       \tikzstyle{exportrepair}=[draw,rectangle,dotted, very thick,white!80!black,text=white!70!black];
      \tikzstyle{repairreaction}=[draw,rectangle,white!40!black,text=white!40!black,dashed];
      \tikzstyle{solreaction}=[draw,rectangle,LimeGreen,text=black];
      \tikzstyle{initial}=[->,>=latex,thick];
      \tikzstyle{bdd}=[->,>=latex,thick];
      \tikzstyle{etiq}=[midway,fill=black!20,scale=0.5];
      \tikzstyle{stc}=[draw, rectangle, white, text=black]

      \draw [black,dotted, rounded corners, very thick] (-0.5,4) rectangle (10.2,9);

      \node[seed] (S) at (0,7) {$S$};
      \node[metabolite] (A) at (3,8) {$A$};
      \node[metabolite] (B) at (3,6) {$B$};
      \node[metabolite] (C) at (6.2,7) {$C$};
      \node[metabolite] (D) at (9.5,7) {$D$};

      \node[reaction] (R1) at (1.5,7) {$r_{1}$};
      \node[reaction, very thick,YellowOrange] (R4) at (7.7,7) {$r_{4}$}; 
      \node[solreaction] (R3) at (4.7,6.4) {$r_{3}$};

      \draw[->,>=latex] (S.east) -- (R1.west);
      \draw[->,>=latex] (R1.east) -- (2.2,7) -- (A.west);
      \draw[->,>=latex] (2.2,7)  -- (B.west);


      \draw[->,>=latex,LimeGreen] (B.east) -- (R3.west);
      \draw[->,>=latex,LimeGreen] (R3.east) -- (C.south west);

      \draw[->,>=latex] (C.east) -- (R4.west);
      \draw[->,>=latex] (R4.east) -- (D.west);

      \node[export] (outD) at (9.5,3) {\ExportReaction};
      \draw[->,>=stealth,white!30!black,dotted, very thick] (D.south) --  (outD.north);

      \node[export] (inS) at (0,10) {$r_{s}$};
      \draw[->,>=stealth,white!30!black,dotted, very thick] (inS) --  (S.north);

  \end{tikzpicture}

      \caption{Topological completion $R_2=\{r_3\}$ satisfies $r_4\in\Activity{t}{G_2}{\{S\}}$ and carries no flux as well, due to accumulation of compound $A$ that contradicts Eq.~\ref{eq:stoichiometric:equation}.\label{gra:union_nf_f2}}
    \end{minipage}
    \begin{minipage}[t]{.32\textwidth}
%
%


    \usetikzlibrary{shapes.misc, positioning}
    \begin{tikzpicture}[scale=0.39]\tiny
      \tikzstyle{metabolite}=[draw,circle,fill=white!80!black];
      \tikzstyle{repairmetabolite}=[draw,white!40!black, circle,fill=white!90!black,text=white!40!black,dashed];
      \tikzstyle{seed}=[draw,circle,fill=BlueGreen!70];
      \tikzstyle{target}=[draw,circle,fill=YellowOrange];
      \tikzstyle{reaction}=[draw,rectangle];
       \tikzstyle{export}=[draw,rectangle,dotted, very thick];
       \tikzstyle{exportrepair}=[draw,rectangle,dotted, very thick,white!80!black,text=white!70!black];
      \tikzstyle{repairreaction}=[draw,rectangle,white!40!black,text=white!40!black,dashed];
      \tikzstyle{solreaction}=[draw,rectangle,LimeGreen,text=black];
      \tikzstyle{initial}=[->,>=latex,thick];
      \tikzstyle{bdd}=[->,>=latex,thick];
      \tikzstyle{etiq}=[midway,fill=black!20,scale=0.5];
      \tikzstyle{stc}=[draw, rectangle, white, text=black]

      \draw [black,dotted, rounded corners, very thick] (-0.5,4) rectangle (10.2,9);

      \node[seed] (S) at (0,7) {$S$};
      \node[metabolite] (A) at (3,8) {$A$};
      \node[metabolite] (B) at (3,6) {$B$};
      \node[metabolite] (C) at (6.2,7) {$C$};
      \node[metabolite] (D) at (9.5,7) {$D$};

      \node[reaction] (R1) at (1.5,7) {$r_{1}$};
      \node[reaction, very thick,YellowOrange] (R4) at (7.7,7) {$r_{4}$}; 
      \node[solreaction] (R2) at (4.7,7.6) {$r_{2}$};
      \node[solreaction] (R3) at (4.7,6.4) {$r_{3}$};

      \draw[->,>=latex] (S.east) -- (R1.west);
      \draw[->,>=latex] (R1.east) -- (2.2,7) -- (A.west);
      \draw[->,>=latex] (2.2,7)  -- (B.west);

      \draw[->,>=latex,LimeGreen] (A.east) -- (R2.west);
      \draw[->,>=latex,LimeGreen] (R2.east) -- (C.north west);

      \draw[->,>=latex,LimeGreen] (B.east) -- (R3.west);
      \draw[->,>=latex,LimeGreen] (R3.east) -- (C.south west);

      \draw[->,>=latex] (C.east) -- (R4.west);
      \draw[->,>=latex] (R4.east) -- (D.west);

      \node[export] (outD) at (9.5,3) {\ExportReaction};
      \draw[->,>=stealth,white!30!black,dotted, very thick] (D.south) --  (outD.north);

      \node[export] (inS) at (0,10) {$r_{s}$};
      \draw[->,>=stealth,white!30!black,dotted, very thick] (inS) --  (S.north);

  \end{tikzpicture}

      \caption{Completion with the union $R_1\cup R_2=\{r_2,r_3\}$. $G=G_1\cup G_2$  satisfies $r_4\in\Activity{h}{G}{\{S\}}$ and thus is flux-balanced.
      \label{gra:union_nf_f}}
    \end{minipage}
\end{figure}

\begin{figure}
    \captionsetup{width=0.3\textwidth}
    \centering
    \begin{minipage}[t]{.32\textwidth}
%
%


    \usetikzlibrary{shapes.misc, positioning}
    \begin{tikzpicture}[scale=0.39]\tiny
      \tikzstyle{metabolite}=[draw,circle,fill=white!80!black];
      \tikzstyle{repairmetabolite}=[draw,white!40!black, circle,fill=white!90!black,text=white!40!black,dashed];
      \tikzstyle{seed}=[draw,circle,fill=BlueGreen!70];
      \tikzstyle{target}=[draw,circle,fill=YellowOrange];
      \tikzstyle{reaction}=[draw,rectangle];
       \tikzstyle{export}=[draw,rectangle,dotted, very thick];
       \tikzstyle{exportrepair}=[draw,rectangle,dotted, very thick,white!80!black,text=white!70!black];
      \tikzstyle{repairreaction}=[draw,rectangle,white!40!black,text=white!40!black,dashed];
      \tikzstyle{solreaction}=[draw,rectangle,LimeGreen,text=black];
      \tikzstyle{initial}=[->,>=latex,thick];
      \tikzstyle{bdd}=[->,>=latex,thick];
      \tikzstyle{etiq}=[midway,fill=black!20,scale=0.5];
      \tikzstyle{stc}=[draw, rectangle, white, text=black]

      \draw [black,dotted, rounded corners, very thick] (-0.5,4) rectangle (10.2,9);

      \node[seed] (S) at (0,7) {$S$};
      \node[metabolite] (A) at (3,8) {$A$};
      \node[metabolite] (B) at (3,6) {$B$};
      \node[metabolite] (C) at (6.2,7) {$C$};
      \node[metabolite] (D) at (9.5,7) {$D$};
      \node[metabolite] (E) at (6.2,5.5) {$E$};

      \node[reaction] (R1) at (1.5,7) {$r_{1}$};
      \node[reaction, very thick,YellowOrange] (R4) at (7.7,7) {$r_{4}$}; 
      \node[solreaction] (R2) at (4.7,7.6) {$r_{2}$};

      \draw[->,>=latex] (S.east) -- (R1.west);
      \draw[->,>=latex] (R1.east) -- (2.2,7) -- (A.west);
      \draw[->,>=latex] (2.2,7)  -- (B.west);

      \draw[->,>=latex,LimeGreen] (A.east) -- (R2.west);
      \draw[->,>=latex,LimeGreen] (R2.east) -- (C.north west);


      \draw[->,>=latex] (C.east) -- (R4.west);
      \draw[->,>=latex] (R4.east) -- (D.west);

      \node[export] (outD) at (9.5,3) {\ExportReaction};
      \draw[->,>=stealth,white!30!black,dotted, very thick] (D.south) --  (outD.north);

      \node[export] (inS) at (0,10) {$r_{s}$};
      \draw[->,>=stealth,white!30!black,dotted, very thick] (inS) --  (S.north);

  \end{tikzpicture}

      \caption{Topological completion $R_1=\{r_2\}$ satisfies $r_4\in\Activity{t}{G_1}{\{S\}}$, but carries no flux, due to accumulation of compound $B$ that contradicts Eq.~\ref{eq:stoichiometric:equation}.\label{gra:union_nf_nf1}}
    \end{minipage}
    \begin{minipage}[t]{.32\textwidth}
%
%


    \usetikzlibrary{shapes.misc, positioning}
    \begin{tikzpicture}[scale=0.39]\tiny
      \tikzstyle{metabolite}=[draw,circle,fill=white!80!black];
      \tikzstyle{repairmetabolite}=[draw,white!40!black, circle,fill=white!90!black,text=white!40!black,dashed];
      \tikzstyle{seed}=[draw,circle,fill=BlueGreen!70];
      \tikzstyle{target}=[draw,circle,fill=YellowOrange];
      \tikzstyle{reaction}=[draw,rectangle];
       \tikzstyle{export}=[draw,rectangle,dotted, very thick];
       \tikzstyle{exportrepair}=[draw,rectangle,dotted, very thick,white!80!black,text=white!70!black];
      \tikzstyle{repairreaction}=[draw,rectangle,white!40!black,text=white!40!black,dashed];
      \tikzstyle{solreaction}=[draw,rectangle,LimeGreen,text=black];
      \tikzstyle{initial}=[->,>=latex,thick];
      \tikzstyle{bdd}=[->,>=latex,thick];
      \tikzstyle{etiq}=[midway,fill=black!20,scale=0.5];
      \tikzstyle{stc}=[draw, rectangle, white, text=black]

      \draw [black,dotted, rounded corners, very thick] (-0.5,4) rectangle (10.2,9);

      \node[seed] (S) at (0,7) {$S$};
      \node[metabolite] (A) at (3,8) {$A$};
      \node[metabolite] (B) at (3,6) {$B$};
      \node[metabolite] (C) at (6.2,7) {$C$};
      \node[metabolite] (D) at (9.5,7) {$D$};
      \node[metabolite] (E) at (6.2,5.5) {$E$};

      \node[reaction] (R1) at (1.5,7) {$r_{1}$};
      \node[reaction, very thick,YellowOrange] (R4) at (7.7,7) {$r_{4}$}; 
      \node[solreaction] (R3) at (4.7,6.4) {$r_{3}$};

      \draw[->,>=latex] (S.east) -- (R1.west);
      \draw[->,>=latex] (R1.east) -- (2.2,7) -- (A.west);
      \draw[->,>=latex] (2.2,7)  -- (B.west);


      \draw[->,>=latex,LimeGreen] (B.east) -- (R3.west);
      \draw[->,>=latex,LimeGreen] (R3.east) -- (C.south west);
      \draw[->,>=latex,LimeGreen] (R3.east) -- (E.north west);

      \draw[->,>=latex] (C.east) -- (R4.west);
      \draw[->,>=latex] (R4.east) -- (D.west);

      \node[export] (outD) at (9.5,3) {\ExportReaction};
      \draw[->,>=stealth,white!30!black,dotted, very thick] (D.south) --  (outD.north);

      \node[export] (inS) at (0,10) {$r_{s}$};
      \draw[->,>=stealth,white!30!black,dotted, very thick] (inS) --  (S.north);

  \end{tikzpicture}

      \caption{Topological completion $R_1=\{r_3\}$ satisfies $r_4\in\Activity{t}{G_2}{\{S\}}$, but carries no flux, due to accumulation of compounds $A$ and $E$ that contradicts Eq.~\ref{eq:stoichiometric:equation}.\label{gra:union_nf_nf2}}
    \end{minipage}
    \begin{minipage}[t]{.32\textwidth}
%
%


    \usetikzlibrary{shapes.misc, positioning}
    \begin{tikzpicture}[scale=0.39]\tiny
      \tikzstyle{metabolite}=[draw,circle,fill=white!80!black];
      \tikzstyle{repairmetabolite}=[draw,white!40!black, circle,fill=white!90!black,text=white!40!black,dashed];
      \tikzstyle{seed}=[draw,circle,fill=BlueGreen!70];
      \tikzstyle{target}=[draw,circle,fill=YellowOrange];
      \tikzstyle{reaction}=[draw,rectangle];
       \tikzstyle{export}=[draw,rectangle,dotted, very thick];
       \tikzstyle{exportrepair}=[draw,rectangle,dotted, very thick,white!80!black,text=white!70!black];
      \tikzstyle{repairreaction}=[draw,rectangle,white!40!black,text=white!40!black,dashed];
      \tikzstyle{solreaction}=[draw,rectangle,LimeGreen,text=black];
      \tikzstyle{initial}=[->,>=latex,thick];
      \tikzstyle{bdd}=[->,>=latex,thick];
      \tikzstyle{etiq}=[midway,fill=black!20,scale=0.5];
      \tikzstyle{stc}=[draw, rectangle, white, text=black]

      \draw [black,dotted, rounded corners, very thick] (-0.5,4) rectangle (10.2,9);

      \node[seed] (S) at (0,7) {$S$};
      \node[metabolite] (A) at (3,8) {$A$};
      \node[metabolite] (B) at (3,6) {$B$};
      \node[metabolite] (C) at (6.2,7) {$C$};
      \node[metabolite] (D) at (9.5,7) {$D$};
      \node[metabolite] (E) at (6.2,5.5) {$E$};

      \node[reaction] (R1) at (1.5,7) {$r_{1}$};
      \node[reaction, very thick,YellowOrange] (R4) at (7.7,7) {$r_{4}$}; 
      \node[solreaction] (R2) at (4.7,7.6) {$r_{2}$};
      \node[solreaction] (R3) at (4.7,6.4) {$r_{3}$};

      \draw[->,>=latex] (S.east) -- (R1.west);
      \draw[->,>=latex] (R1.east) -- (2.2,7) -- (A.west);
      \draw[->,>=latex] (2.2,7)  -- (B.west);

      \draw[->,>=latex,LimeGreen] (A.east) -- (R2.west);
      \draw[->,>=latex,LimeGreen] (R2.east) -- (C.north west);

      \draw[->,>=latex,LimeGreen] (B.east) -- (R3.west);
      \draw[->,>=latex,LimeGreen] (R3.east) -- (C.south west);
      \draw[->,>=latex,LimeGreen] (R3.east) -- (E.north west);

      \draw[->,>=latex] (C.east) -- (R4.west);
      \draw[->,>=latex] (R4.east) -- (D.west);

      \node[export] (outD) at (9.5,3) {\ExportReaction};
      \draw[->,>=stealth,white!30!black,dotted, very thick] (D.south) --  (outD.north);

      \node[export] (inS) at (0,10) {$r_{s}$};
      \draw[->,>=stealth,white!30!black,dotted, very thick] (inS) --  (S.north);

  \end{tikzpicture}

      \caption{Completion with the union $R_1\cup R_2=\{r_2,r_3\}$. $G=G_1\cup G_2$ satisfies $r_4\in\Activity{t}{G}{\{S\}}$, but contradicts minimality and carries no flux $r_4\not \in\Activity{s}{G}{\{S\}}$, due to accumulation of compound $E$ that contradicts Eq.~\ref{eq:stoichiometric:equation}.\label{gra:union_nf_nf}}
    \end{minipage}
\end{figure}


\section{Answer Set Programming with Linear Constraints}\label{sec:background}

For encoding our hybrid problem,
we rely upon the theory reasoning capacities of the ASP system \clingo\ that allows us to extend ASP with linear constraints over reals
(as addressed in Linear Programming).
We confine ourselves below to features relevant to our application and refer the interested reader for details to~\citep{gekakaosscwa16a}.

As usual, a \emph{logic program} consists of \emph{rules} of the form
\begin{lstlisting}[mathescape=true,numbers=none]
   a$_0$ :- a$_1$,...,a$_m$,not a$_{m+1}$,...,not a$_n$
\end{lstlisting}
where each \lstinline[mathescape=true]{a$_i$} is either
a \emph{(regular) atom} of form \lstinline[mathescape=true]{p(t$_1$,...,t$_k$)}
where all \lstinline[mathescape=true]{t$_i$} are terms
or
a \emph{linear constraint atom} of form%
\footnote{In \clingo, theory atoms are preceded by `\texttt{\&}'.}
`\lstinline[mathescape=true]@&sum{w$_1$*x$_1$;$\dots$;w$_l$*x$_l$} <= k@'
that stands for the linear constraint
\(
w_1\cdot x_1+\dots+w_l\cdot x_l\leq k
\).
All \lstinline[mathescape=true]{w$_i$} and \lstinline[mathescape=true]{k} are finite sequences of digits with at most one dot%
\footnote{In the input language of \clingo, such sequences must be quoted to avoid clashes.}
and represent real-valued coefficients $w_i$ and $k$.
Similarly all \lstinline[mathescape=true]{x$_i$} stand for the real-valued variables $x_i$.
As usual, \lstinline[mathescape=true]{not} denotes (default) \emph{negation}.
A rule is called a \emph{fact} if $n=0$.

Semantically, a logic program induces a set of \emph{stable models},
being distinguished models of the program determined by stable models semantics~\citep{gellif91a}.
Such a stable model $X$ is an \emph{LC-stable model} of a logic program $P$,%
\footnote{This corresponds to the definition of $T$-stable models using a \emph{strict} interpretation of theory atoms~\citep{gekakaosscwa16a},
  and letting $T$ be the theory of linear constraints over reals.}
if there is an assignment of reals to all real-valued variables occurring in $P$ that
(i)     satisfies all linear constraints associated with linear constraint atoms in $P$ being     in $X$
and
(ii) falsifies all linear constraints associated with linear constraint atoms in $P$ being not in $X$.
For instance, the (non-ground) logic program containing the fact
`\lstinline[mathescape=true]{a("1.5").}'
along with the rule
`\lstinline[mathescape=true]@&sum{R*x} <= 7 :- a(R).@'
has the stable model
\par
\lstinline[mathescape=true]@$\{$a("1.5")$,\;$&sum{"1.5"*x}<=7$\}$@.
\\
This model is LC-stable since there is an assignment,
e.g.\ $\{x\mapsto 4.2\}$,
that satisfies the associated linear constraint `$1.5*x\leq 7$'.
We regard the stable model along with a satisfying real-valued assignment as a solution to a logic program containing linear constraint atoms.
\review{For a more detailed introduction of ASP extended with linear constraints, illustrated with more complex examples, we refer the interested reader to~\citep{jakaosscscwa17a}.}

To ease the use of ASP in practice,
several extensions have been developed.
First of all, rules with variables are viewed as shorthands for the set of their ground instances.
Further language constructs include
\emph{conditional literals} and \emph{cardinality constraints} \citep{siniso02a}.
The former are of the form
\lstinline[mathescape=true]{a:b$_1$,...,b$_m$},
the latter can be written as
\lstinline[mathescape=true]+s{d$_1$;...;d$_n$}t+,
where \lstinline{a} and \lstinline[mathescape=true]{b$_i$} are possibly default-negated (regular) literals  
and each \lstinline[mathescape=true]{d$_j$} is a conditional literal; 
\lstinline{s} and \lstinline{t} provide optional lower and upper bounds on the number of satisfied literals in the cardinality constraint.
We refer to \lstinline[mathescape=true]{b$_1$,...,b$_m$} as a \emph{condition}.
The practical value of both constructs becomes apparent when used with variables.
For instance, a conditional literal like
\lstinline[mathescape=true]{a(X):b(X)}
in a rule's antecedent expands to the conjunction of all instances of \lstinline{a(X)} for which the corresponding instance of \lstinline{b(X)} holds.
Similarly,
\lstinline[mathescape=true]+2{a(X):b(X)}4+
is true whenever at least two and at most four instances of \lstinline{a(X)} (subject to \lstinline{b(X)}) are true.
Finally, objective functions minimizing the sum of weights $w_i$ subject to condition $c_i$ are expressed as
\lstinline[mathescape=true]!#minimize{$w_1$:$c_1$;$\dots$;$w_n$:$c_n$}!.

In the same way,
the syntax of linear constraints offers several convenience features.
As above,
elements in linear constraint atoms can be conditioned,
viz.\par
`\lstinline[mathescape=true]@&sum{w$_1$*x$_1$:c$_1$;...;w$_l$*x$_l$:c$_n$} <= k@'
\\
where each \lstinline[mathescape=true]{c$_i$} is a condition.
Moreover, the theory language for linear constraints offers a domain declaration for real variables,
`\lstinline[mathescape=true]@&dom{lb..ub} = x@'
expressing that all values of \texttt{x} must lie between \texttt{lb} and \texttt{ub}.
And finally the maximization (or minimization) of an objective function can be expressed with
\lstinline[mathescape=true]@&maximize{w$_1$*x$_1$:c$_1$;...;w$_l$*x$_l$:c$_n$}@
(by \texttt{minimize}).
The full theory grammar for linear constraints over reals is available at~\url{https://potassco.org}.


\section{Solving Hybrid Metabolic Network Completion}
\label{sec:approach}

In this section, we present our hybrid approach to metabolic network completion.
We start with a factual representation of problem instances.
A metabolic network $G$ with a typing function $t: M\cup R\rightarrow\{\texttt{d,r,s,t}\}$,
indicating the origin of the respective entities,
is represented as follows:
\begin{align*}
F(G,t) = & \phantom{\cup\;}\;\{\texttt{metabolite($m$,$t(m)$)}\mid m\in M\} \\
         &          \cup\;   \{\texttt{reaction($r$,$t(r)$)}\mid r\in R\}  \\
         &          \cup\;   \{\texttt{bounds($r$,$lb_r$,$ub_r$)$\mid r\in R$}\} \;\cup\; \{\texttt{objective($r$,$t(r)$)$\mid r\in R$}\}\\
         &          \cup\;   \{\texttt{reversible(r)}\mid r\in R, \Reactants{r}\cap\Products{r}\neq\emptyset\} \\
         &          \cup\;   \{\texttt{rct($m$,$\StoichiometricFunction(m,r)$,$r$,$t(r)$)$\mid r\in R, m\in\Reactants{r}$}\} \\
         &          \cup\;   \{\texttt{prd($m$,$\StoichiometricFunction(r,m)$,$r$,$t(r)$)$\mid r\in R, m\in\Products{r}$}\}
\end{align*}
While most predicates should be self-explanatory,
we mention that \texttt{reversible} identifies bidirectional reactions.
Only one direction is explicitly represented in our fact format.
The four types \texttt{d}, \texttt{r}, \texttt{s}, and \texttt{t} tell us whether an entity stems from the
\textbf{d}raft or \textbf{r}eference network, or belongs to the \textbf{s}eeds or \textbf{t}argets.

In a metabolic network completion problem,
we consider
a draft network  $G=(R\cup M,E,\StoichiometricFunction)$,
a set $S$ of seed compounds, 
a set $R_{T}$ of target reactions,
and a reference network $G'=(R'\cup M',E',\StoichiometricFunction')$.
An instance of this problem is represented by the set of facts
\(
F(G,t)\cup F(G',t')
\).
In it, a key role is played by the typing functions that differentiate the various components:
\[
  t(n) =
  \left\{
    \begin{array}{ll}
    \texttt{d}, & \text{if } n\in (M\setminus (T\cup S))\cup (R\setminus(R_{\strseed}\cup R_T)) \\
    \texttt{s}, & \text{if } n\in S\cup R_{\strseed} \\
    \texttt{t}, & \text{if } n\in T\cup R_T
  \end{array}
  \right.
  \quad\text{ and }\quad
  t'(n) = \texttt{r},
\]
where
\(
T=\{m\in\Reactants{r}\mid r\in R_T\}
\)
is the set of target compounds and
\(
R_{\strseed}=\{r\in R\mid m\in \strseed(G), m\in\Products{r}\}
\)
is the set of reactions related to boundary seeds.

Our encoding of hybrid metabolic network completion is given in Listing~\ref{lst:encoding}.
\begin{lstlisting}[float=t,floatplacement=t,numbers=left,numberblanklines=false,basicstyle=\ttfamily\scriptsize,firstline=1,lastline=27,caption={Encoding of hybrid metabolic network completion},label=lst:encoding,belowskip=-2em]
edge(R,M,N,T) :- reaction(R,T), rct(M,_,R,T), prd(N,_,R,T).
edge(R,M,N,T) :- reaction(R,T), rct(N,_,R,T), prd(M,_,R,T), reversible(R).

scope(M,d) :- metabolite(M,s).
scope(M,d) :- edge(R,_,M,T), T!=r, scope(N,d):edge(R,N,_,T'), N!=M, T'!=r.

scope(M,x) :- scope(M,d). 
scope(M,x) :- edge(R,_,M,_), scope(N,x):edge(R,N,_,_), N!=M.

{ completion(R) : edge(R,M,N,r), scope(N,x), scope(M,x) }.

scope(M,c) :- scope(M,d). 
scope(M,c) :- edge(R,_,M,T), T!=r, scope(N,c):edge(R,N,_,T'), T'!=r, N!=M.
scope(M,c) :- completion(R), edge(R,_,M,r), scope(N,c):edge(R,N,_,r), N!=M.

:- metabolite(M,t), not scope(M,c). 

&dom{L..U} = R :- bounds(R,L,U).

&sum{ IS*IR : prd(M,IS,IR,T), T!=r;  IS'*IR' : prd(M,IS',IR',r), completion(IR');
     -OS*OR : rct(M,OS,OR,T), T!=r; -OS'*OR' : rct(M,OS',OR',r), completion(OR')
    } = "0" :- metabolite(M,_).

&sum{ R } > "0" :- reaction(R,t).

&maximize{   R : objective(R,t) }. 
#minimize{ 1,R : completion(R)  }.
\end{lstlisting}
Roughly,
the first 10 lines lead to a set of candidate reactions for completing the draft network.
Their topological validity is checked in lines~12--16 with regular ASP,
the stoichiometric one in lines~18--24 in terms of linear constraints.
(Lines~1--16 constitute a revision of the encoding in~\citep{schthi09a}.)
The last two lines pose a hybrid optimization problem,
first minimizing the size of the completion and then
maximizing the flux of the target reactions.%

In more detail,
we begin by defining the auxiliary predicate \texttt{edge}/4 representing directed edges between compounds connected by a reaction.
With it,
we calculate in Line~4 and~5 the scope $\Sigma_{G}(S)$ of the \textbf{d}raft network $G$ from the seed compounds in $S$;
it is captured by all instances of \texttt{scope(M,d)}.
This scope is then e\textbf{x}tended in Line 7/8 via the reference network $G'$ to delineate all possibly producible compounds.
We draw on this in Line~10 when choosing the reactions $R''$ of the completion (cf.\ Section~\ref{sec:problem})
by restricting their choice to reactions from the reference network whose reactants are producible.
This amounts to a topological search space reduction.

The reactions in $R''$ are then used in lines~12--14 to compute the scope $\Sigma_{G''}(S)$ of the \textbf{c}ompleted network.
And $R''$ constitutes a topologically valid completion if all targets in $T$ are producible by the expanded draft network $G''$:
Line~16 checks whether $T\subseteq\Sigma_{G''}(S)$ holds, which is equivalent to $R_T\subseteq\Activity{t}{G''}{S}$.
Similarly, $R''$ is checked for stoichiometric validity in lines~18--24.
For simplicity, we associate reactions with their rate and let their identifiers take real values.
Accordingly, Line~18 accounts for \eqref{eq:stoichiometric:bounds} by imposing lower and upper bounds on each reaction rate.
The mass-balance equation \eqref{eq:stoichiometric:equation} is enforced for each metabolite \texttt{M} in lines~20--22;
it checks whether the sum of products of stoichiometric coefficients and reaction rates equals zero,
viz.\ \texttt{IS*IR}, \texttt{-OS*OR}, \texttt{IS'*IR'}, and \texttt{-OS'*OR'}.
Reactions \texttt{IR}, \texttt{OR} and \texttt{IR'}, \texttt{OR'} belong to the draft and reference network, respectively,
and correspond to $R\cup R''$.
Finally, by enforcing $r_T>0$ for $r_T\in R_T$ in Line~24,
we make sure that $R_T\subseteq\Activity{s}{G''}{S}$.

In all, our encoding ensures that the set $R''$ of reactions chosen in Line~10 induces an augmented network $G''$
in which all targets are activated both topologically as well as stoichiometrically,
and is optimal wrt the hybrid optimization criteria.


\section{System and Experiments}\label{sec:experiments}
\label{sec:sysandexp}

In this section, we introduce \fluto, our new system for hybrid metabolic network completion, and empirically evaluate its performance.
The system relies on the hybrid encoding described in Section~\ref{sec:approach}
along with the hybrid solving capacities of \clingo~\citep{gekakaosscwa16a} for implementing the combination of ASP and LP.
We use \clingo~5.2.0 
incorporating as LP solvers either \cplex~12.7.0.0 or \lpsolve~5.5.2.5 via their respective \python\ interfaces.
We describe the details of the underlying solving techniques in a separate paper and focus below on application-specific aspects.

The output of \fluto\ consists of two parts.
First, the completion $R''$, given by instances of predicate \texttt{completion}, and
second, an assignment of floats to (metabolic flux variables $v_r$ for) all $r\in R\cup R''$.
In our example, we get
\begin{align*}
R''=\{\texttt{completion}(r_6), \texttt{completion}(r_8), \texttt{completion}(r_9)\}
\\
\text{ and } \{\SeedReaction=49999.5, r_9=49999.5, r_3=49999.5, r_2=49999.5, \\
 \ExportReaction=99999.0, r_6=49999.5, r_5=49999.5, r_4=49999.5\}.
\end{align*}

Variables assigned $0$ are omitted.
Note the flux value $r_8=0$ even though $r_8\in R''$.
This is to avoid the self-activation of cycle $C$, $D$ and $E$.
By choosing $r_8$, we ensure that the cycle has been externally initiated at some point
but activation of $r_8$ is not necessary at the current steady state.

We analyze
(i)   the impact of different system configurations
(ii)  the quality of \fluto's approach to metabolic network completion, and
(iii) compare the quality of \fluto's solutions with other approaches.
To have a realistic setting,
we use degradations of a functioning metabolic network of \textit{Escherichia coli}~\citep{Reed2003} comprising 1075 reactions.
The network was randomly degraded by 10, 20, 30 and 40 percent,
creating 10 networks for each degradation
by removing reactions until the target reactions were inactive according to \emph{Flux Variability Analysis}~\citep{Becker2007}.
90 target reactions with varied reactants were randomly chosen for each network, yielding 3600 problem instances in total ~\citep{Prigent2017}.
The reference network consists of reactions of the original metabolic network.

We ran each benchmark on a Xeon E5520 2.4 GHz processor under Linux limiting RAM to 20~GB.
At first,
we investigate two alternative optimization strategies for computing completions of minimum size.
The first one, \emph{branch-and-bound}~(\bb), iteratively produces solutions of better quality until the optimum is found and the other,
\emph{unsatisfiable core}~(\usc), relies on successively identifying and relaxing unsatisfiable cores until an optimal solution is obtained.
Note that we are not only interested in optimal solutions
but if unavailable also solutions activating target reactions without trivially restoring the whole reference network.
In \clingo, \bb\ naturally produces these solutions in contrast to \usc.
Therefore, we use \usc\ with stratification~\citep{anbole13a}, which provides at least some suboptimal solutions.

\subsection{System configurations}
\label{sec:sysconf}
\newcommand{\shade}[2]
{%
    \cellcolor{black!\xinttheiexpr 80*#1/#2-30\relax}%
       {#1}%
}%

    \begin{table}[t]
        \begin{center}
            \begin{tabular}{r|*{5}{c}}
				\backslashbox{\core{}}{\prop{}} & 0 & 25 & 50 & 75 & 100\\\hline
0 & \shade{383.20}{600}(130) & \shade{388.51}{600}(134) & \shade{384.46}{600}(133) & \shade{388.45}{600}(137) & \shade{398.21}{600}(134)\\
25 & \shade{385.05}{600}(132) & \shade{385.95}{600}(133) & \shade{391.84}{600}(131) & \shade{382.29}{600}(137) & \shade{401.73}{600}(134)\\
50 & \shade{383.95}{600}(131) & \shade{377.51}{600}(123) & \shade{385.46}{600}(132) & \shade{391.05}{600}(137) & \shade{399.88}{600}(141)\\
75 & \shade{358.77}{600}(129) & \shade{360.54}{600}(127) & \shade{356.89}{600}(131) & \shade{390.69}{600}(137) & \shade{399.36}{600}(134)\\
100 & \shade{376.03}{600}(133) & \shade{370.75}{600}(132) & \shade{375.77}{600}(133) & \shade{389.77}{600}(139) & \shade{401.18}{600}(139)\\

            \end{tabular}
        \end{center}
        \caption{Comparison of propagation and core minimization heuristics for \bb. \label{tab:pchbb}}
    \end{table}

    \begin{table}[t]
        \begin{center}
            \begin{tabular}{r|*{5}{c}}
				\backslashbox{\core{}}{\prop{}} & 0 & 25 & 50 & 75 & 100\\\hline
0 & \shade{297.38}{600}(102) & \shade{296.39}{600}(102) & \shade{296.48}{600}(103) & \shade{299.14}{600}(105) & \shade{475.12}{600}(200)\\
25 & \shade{297.29}{600}(101) & \shade{293.69}{600}(100) & \shade{297.09}{600}(102) & \shade{293.43}{600}(101) & \shade{478.39}{600}(202)\\
50 & \shade{292.65}{600}(102) & \shade{296.43}{600}(102) & \shade{294.4}{600}(103) & \shade{295.48}{600}(102) & \shade{477.67}{600}(200)\\
75 & \shade{331.72}{600}(127) & \shade{336.34}{600}(129) & \shade{331.17}{600}(127) & \shade{294.17}{600}(103) & \shade{476.16}{600}(202)\\
100 & \shade{308.88}{600}(108) & \shade{309.47}{600}(107) & \shade{324.9}{600}(122) & \shade{489.97}{600}(214) & \shade{476.17}{600}(201)\\

            \end{tabular}
        \end{center}
        \caption{Comparison of propagation and core minimization heuristics for \usc. \label{tab:pchusc}}
    \end{table}

The configuration space of \fluto\ is huge.
In addition to its own parameters, the ones of \clingo\ and the respective LP solver amplify the number of options.
We thus concentrate on distinguished features revealing an impact in our experiments.

The first focus are two system options controlling the hybrid solving nature of \fluto.
First, \prop{$n$} controls the frequency of LP propagation:
the consistency of linear constraints is only checked if $n\%$ of atoms are decided.
Second,
the \fluto\ option \core{$n$} invokes the irreducible inconsistent set algorithm~\citep{ostsch12a} whenever $n\%$ of atoms are decided.
This algorithm extracts a minimal set of conflicting linear constraints for a given conflict.
Note that the second parameter depends on the first one,
since conflict analysis may only be invoked if the LP solver found an inconsistency.

The \default{} is to use \core{100}, \prop{0}, and use LP solver \cplex\footnote{We do not present results of \lpsolve\ since it produced inferior results.}.
This allows us to detect conflicts among the linear constraints as soon as possible and only perform expensive conflict analysis on the full assignment.

To get an overview, we conducted a preliminary experiment
using \bb\ and \usc\ with \fluto's default configuration on the 10, 20, and 30 percent degraded networks,
2700 instances in total,
limiting execution time to 20 minutes.
For our performance experiments,
we selected at random three networks with at least one instance
for which \bb\ and \usc\ could find the optimum in 100 to 600 seconds.
With the resulting 270 medium to hard instances,
we examined the cross product of values $n\in\{0,25,50,75,100\}$ for \core{n} and \prop{n}, respectively, limiting time to 600 seconds.

Table~\ref{tab:pchbb} and Table~\ref{tab:pchusc} display the results using \bb\ and \usc\ respectively.
The columns increase the value for \prop{n} and the rows for \core{n} in steps of 25,
i.e., LP propagation becomes less frequent from left to right,
and conflict minimization from top to bottom.
The first value in each cell is the average runtime in seconds and the value in brackets shows the number of timeouts.
The shade of the cells depends on the average runtime,
i.e., the darker the cell, the less performant the combination of propagation and conflict minimization heuristics.

Table~\ref{tab:pchbb} shows that propagation and conflict minimization heuristics have an overall small impact on the performance of \bb\ optimization.
Since \bb\ relies on iterating solutions and learns weaker constraints,
only pertaining to the best known bound, while optimizing,
the improvement step is less constraint compared to \usc.
Due to this, conflicts are more likely to appear later on in the optimization process allowing for less impact of frequent LP propagation and conflict minimization
Nevertheless, we see a slight performance improvement of propagating and conflict minimizing for every partial assignment (\prop{0}, \core{0})
compared to only on full assignments (\prop{100}, \core{100}).
To prove the optimum, the solver is still required to cover the whole search space.
For this purpose, early pruning and conflict minimization may be effective.
Furthermore, we see the best average runtime in the area \prop{0-50} at \core{75}.
That indicates a good tradeoff between the better quality conflicts which prune the search effectively
and the overhead of the costly conflict minimization.
There is no clear best configuration,
 but \prop{25} and \core{75} shows the best tradeoff between average runtime and number of timeouts.

\usc\ on the other hand (Table~\ref{tab:pchusc}), clearly benefits from early propagation and conflict minimization.
The area \prop{0-75} and \core{0-50} has the lowest average runtime and number of timeouts,
best among them \prop{25} and \core{25}
with the lowest timeouts and average runtime that is not significantly different from the best value.
\usc\ aims at quickly identifying unsatisfiable partial assignments
and learning structural constraints building upon each other,
which is enhanced by frequent conflict detection and minimization.
Disabling LP propagation on partial assignments with \usc\ leads to the overall worst performance
and we also see deterioration with \core{75} and \core{100} in the interval \prop{0-50}.
Overall, \usc\ is more effective than \bb\ for the instances and we see a benefit in early LP propagation and conflict minimization as well as in fine-tuning the heuristics at which point both are applied.

    \begin{table}[t]
        \begin{center}
            \begin{tabular}{r|cccccccccccc}
            	 & \multicolumn{2}{c}{\textsc{FR}}& \multicolumn{2}{c}{\textsc{JP}}& \multicolumn{2}{c}{\textsc{TW}}& \multicolumn{2}{c}{\textsc{TR}}& \multicolumn{2}{c}{\textsc{CR}}& \multicolumn{2}{c}{\textsc{HD}}\\
              & \T & \TO & \T & \TO & \T & \TO & \T & \TO & \T & \TO & \T & \TO\\\hline
             \bb  & 400.41 & 154 & 389.68 & 147 & \textbf{360.54} & 127 & 409.33 & 141 & 362.74 & \textbf{120} & 434.54 & 160\\
             \usc & 227.38 &  78 & 293.96 & 100 & 316.54 & 107 & 293.54 & 102 & \textbf{221.84} &  \textbf{74} & 297.32 & 104\\
            \end{tabular}
        \end{center}
        \caption{Comparison of \clingo's portfolio configurations for \bb\ and \usc. \label{tab:search}}
    \end{table}

Now, we focus on the portfolio configurations of \clingo.
Those configurations were crafted by experts to enhance the solving performance of problems with certain attributes.
\review{To examine their impact, we take the best result for \bb\ (\prop{25} and \core{75}) and \usc\ (\prop{25} and \core{25}), }
and employ the following \clingo\ options:
\begin{description}
	\item[\emph{\textsc{FR}}]
      Refers to {\clingo}'s configuration \emph{frumpy} that uses more conservative defaults.
    \item[\emph{\textsc{JP}}]
      Refers to {\clingo}'s configuration \emph{jumpy} that uses more aggressive defaults.
    \item[\emph{\textsc{TW}}]
      Refers to {\clingo}'s configuration \emph{tweety} that is geared toward typical ASP problems.
    \item[\emph{\textsc{TR}}]
      Refers to {\clingo}'s configuration \emph{trendy} that is geared toward industrial problems.
	\item[\emph{\textsc{CR}}]
      Refers to {\clingo}'s configuration \emph{crafty} that is geared towards crafted problems.
    \item[\emph{\textsc{HD}}]
      Refers to {\clingo}'s configuration \emph{handy} that is geared towards larger problems.
\end{description}
For more information on {\clingo}'s configurations, see ~\citep{gekakarosc15a}.

Table~\ref{tab:search} shows the average runtime in seconds (\T) and number of timeouts (\TO)
for all six configurations using \bb\ and \usc\ on the same 270 instances.
Even though \textsc{CR} has slightly higher average runtime for \bb\ compared to \textsc{TW},
it is the overall best configuration.
This configuration is geared towards problems with an inherent structure
compared to randomly generated benchmarks
which fits with the metabolic network completion problem at hand
since the data is taken from an existing bacteria.
Interestingly, \bb\ performs worse under more specific configurations
and favors moderate once like \textsc{TW} and \textsc{CR}.
This might be due to the changing nature of improvement steps as the optimization process goes on
from finding any random solutions to an unsatisfiability proof in the end.
\usc\ on the other hand, benefits from a more structural heuristics in \textsc{CR} and more conservative defaults in \textsc{FR}
which allow the solver to explore and collect conflicts instead of frequently restarting and forgetting.

\subsection{Solution quality}

\begin{table}[t]%
\newcommand{\mc}[3]{\multicolumn{#1}{#2}{#3}}
\centering
\begin{tabular}{r|rr|rr|rr||r} 
 & \mc{2}{c|}{\f(\bb)}  & \mc{2}{c|}{\f(\usc)} & \mc{2}{c}{\f(\bb+\usc)} & \f(\bb+\usc) \\
\degradation & \sols & \opts & \sols & \opts & \sols & \opts & \verified \\
10\% (900) & \textbf{900} & \textbf{900} & 892 & 892 & 900 & 900 & 900 \\ 
20\% (900) & \textbf{830} & 669 & 793 & \textbf{769} & 867 & 814 & 867 \\
30\% (900) & \textbf{718} & 88 & 461 & \textbf{344} & 780 & 382 & 780 \\\hline
all (2700) & \textbf{2448} & 1657 & 2146 & \textbf{2005} & 2547 & 2096 & 2547 \\
\end{tabular}
\caption{Comparison of qualitative results.\label{tab:quality}}
\end{table}

\begin{table}[t]%
\newcommand{\mc}[3]{\multicolumn{#1}{#2}{#3}}
\centering
\begin{tabular}{r|rr||rrr}
 & \mc{2}{c||}{\f(\vbs)} & \mc{3}{c}{\verified}\\
\degradation & \sols & \opts & \f(\vbs)\\\hline
10\% (900) & 900 & 900 & 900\\
20\% (900) & 896 & 855 & 896\\
30\% (900) & 848 & 575 & 848\\
40\% (900) & 681 & 68  & 681\\\hline
all (3600) & 3325 & 2398 & 3325
\end{tabular}
\caption{Results using best system options.\label{tab:final}}
\end{table}

Now, we examine the quality of the solutions provided by \fluto.
Table~\ref{tab:quality} gives the number of solutions~(\sols) and optima~(\opts) obtained by \fluto{}~(\f) in its default setting
within 20 minutes
for \bb, \usc\ and the best of both~(\bb+\usc),
individually for each \textsc{degradation} and over\textbf{all}.
\review{The default setting for \fluto\ includes the default configurations for \clingo\ and \cplex.}
The data was obtained in our preliminary experiment using networks with 10, 20, and 30 percent degradation.
For 94.3\% of the instances \fluto(\bb+\usc) found a solution within the time limit and 82.3\% of them were optimal.
We observe that \bb\ provides overall more useful solutions but \usc\ acquires more optima,
which was to be expected by the nature of the optimization techniques.
Additionally, each technique finds solutions to problem instances where the other exceeds the time limit,
underlining the merit of using both in tandem.
Column~\verified\ shows the quality of solutions provided by \fluto.
Each obtained best solution was checked with \cobrapy~0.3.2~\citep{Ebrahim2013},
a renowned system implementing an FBA-based gold standard (for verification only).
All solutions found by \fluto\ could be verified by \cobrapy.
In detail, \fluto\ found a smallest set of reactions completing the draft network for 77.6\%,
a suboptimal solution for 16.7\%,
and no solution for 5.6\% of the problem instances.

Finally, we change the system configuration and examine how \fluto\ scales on harder instances.
To this end, we use the best configurations from Section~\ref{sec:sysconf},
\prop{25}, \core{75} and \textsc{CR} for \bb, and \prop{25}, \core{25} and \textsc{CR} for \usc,
and rerun the experiment on all 3600 instances.
The results are shown in (Table~\ref{tab:final}).
\f(\vbs) denotes the virtual best results, meaning for each problem instance the best known solution among the two configurations was verified.
For 20\% and 30\% degradation, we obtain additional 29 and 68 solutions and 41 and 193 optima, respectively.
Overall, we find solutions for 92.4\% out of the 3600 instances and 72.1\% of them are optimal.
The number of solutions decreases slightly and the number of optima more drastically with higher degradation.
The results show that \fluto\ is capable of finding correct completions for even highly degraded networks for most of the instances in reasonable time.

\subsection{Comparison to other approaches}

\begin{table}[t]%
\newcommand{\mc}[3]{\multicolumn{#1}{#2}{#3}}
\centering
\begin{tabular}{r|rrr|rrr}
                        & \mc{3}{c|}{\fluto} & \mc{3}{c}{\meneco} \\ 
                        & min & average & max & min & average & max \\ \hline
solutions per instance  & 1   & 2.24    & 12  & 1   & 1.88    & 6 \\
reactions per solution  & 1   & 6.66    & 9   & 1   & 6.24    & 9 \\ \hline
verified solutions     & \mc{3}{r|}{100\%} & \mc{3}{r}{73.39\%} \\
instances with only verified solutions & \mc{3}{r|}{100\%} & \mc{3}{r}{72.94\%} \\
instances without verified solutions   & \mc{3}{r|}{0\%}   & \mc{3}{r}{26.61\%} \\
instance with some verified solutions & \mc{3}{r|}{0\%}   & \mc{3}{r}{0.45\%} \\
\end{tabular}
\caption{Comparison of \fluto\ and \meneco\ solutions for 10 percent degraded networks.\label{tab:enumeration}}
\end{table}

\begin{table}[t]%
\newcommand{\mc}[3]{\multicolumn{#1}{#2}{#3}}
\centering
\begin{tabular}{r|r|r|r}
& \fluto & \meneco & \gapfill \\ \hline
verified union                                 & 100\% & 73.39\% & 6.20\% \\
verified union of verified solutions           & 100\% & 72.94\% & NA \\
verified union of unverified solutions         & 0\%   & 0.00\%  & NA\\
verified union of partially verified solutions & 0\%   & 0.45\%  & NA \\
\end{tabular}
\caption{Comparison of \fluto, \meneco\ and \gapfill\ unions for 10 percent degraded networks.\label{tab:union}}
\end{table}

We compare the quality of \fluto\ with \meneco~1.4.3~\citep{Prigent2017} and \gapfill%
\footnote{Update of 2011-09-23 see http://www.maranasgroup.com/software.htm }~\citep{SatishKumar2007}.
\footnote{The results for \meneco\ and \gapfill\ are taken from previous work~\citep{Prigent2017},
where they were run to completion with \emph{no} time limit.}
Both \meneco\ and \gapfill\ are systems for metabolic network completion.
While \meneco\ pursues the topological approach,
\gapfill\ applies the relaxed stoichiometric variant using Inequation~\eqref{eq:stoichiometric:equation:relaxed}.
We performed an enumeration of all minimal solutions to the completion problem
under the topological (\meneco), the relaxed stoichiometric (\gapfill),
and hybrid (\fluto) activation semantics for the 10 percent degraded networks of the benchmark set (900 instances to be completed).

First, we compare the quality of individual solutions of \fluto\ and \meneco.
\footnote{There was no data available for the individual solutions of \gapfill.}
Results are displayed in Table~\ref{tab:enumeration}.
The first two rows give the minimum, average and maximum number of solutions per instance,
and reactions per solution, respectively, for \fluto\ and \meneco.
While \fluto\ finds 19\% more solutions on average and twice as many maximum solutions per instance compared to \meneco,
the numbers of reactions in minimal solutions of both tools are similar.
The next four rows pertain to the solution quality as established by \cobrapy.
First, what percent solutions over all instances could be verified,
second, what percent of instances had verified solutions exclusively,
third, how many instances had no verified solutions at all,
and finally, percent of instances where only a portion of solutions could be verified.
All of \fluto's solutions could be verified,
compared to the 72.04\% of \meneco\ across all solutions and 72.94\% of instances that were correctly solved.
Interestingly, \meneco\ achieves hybrid activation in some but not all solutions for 0.45\% (4) of the instances.
\fluto\ does not only improve upon the quality of \meneco,
but also provides more solution per instances without increasing the number of relevant reactions significantly.

To empirically evaluate the properties established in Section~\ref{sec:union},
and be able to compare to \gapfill, for which only the union of reactions was available,
we examine the union of minimal solutions provided by all three systems
and present the results in Table~\ref{tab:union}.
The four rows show,
first, for what percent of instances the union of solutions could be verified,
second, how many instances had only verified solutions and their union was also verified,
third, the percentage of instances where the union of solutions displayed activation of the target reactions even though all individual solutions did not provide that,
and forth, instances where the solutions were partly verifiable and their union could also be verified.
While again 100\% of \fluto's solutions could be verified, only 73.3\% and 6.2\% are obtained for \meneco\ and \gapfill, respectively, for 10 percent degraded networks.
As reflected by the results, the ignorance of \meneco\ regarding stoichiometry leads to possibly unbalanced networks.
Still, the union of solutions provided a useful set of reactions in almost three quarters of the instances, showing merit in the topological approximation of the metabolic network completion problem.
On the other hand, the simplified view of \gapfill\ in terms of stoichiometry misguides the search for possible completions and eventually leads to unbalanced networks even in the union.
Moreover, \gapfill's ignorance of network topology results in self-activated cycles.
By exploiting both topology and stoichiometry,
\fluto\ avoids such cycles while still satisfying the stoichiometric activation criteria.
The results support the observations made in Section~\ref{sec:union}.
For both \fluto\ and \meneco\, all instances, for which the complete solution set could be verified,
the union is also verifiable,
as well as all unions for instances where \meneco\ established hybrid activation for a fraction of solutions.


\section{Discussion}\label{sec:discussion} 

We presented the first hybrid approach to metabolic network completion
by combining topological and stoichiometric constraints in a uniform setting.
To this end,
we elaborated a formal framework capturing different semantics for the activation of reactions.
Based upon these formal foundations, we developed a hybrid ASP encoding reconciling
disparate approaches to network completion.
The resulting system, \fluto, thus combines the advantages of both approaches and 
yields greatly superior results compared to purely quantitative or qualitative existing systems.
Our experiments show that \fluto\ scales to more highly degraded networks 
and produces useful solutions in reasonable time. %
In fact, all of \fluto's solutions passed the biological gold standard.
The exploitation of the network's topology guides the solver to more likely completion candidates,
and furthermore avoids self-activated cycles, as obtained in FBA-based approaches.
Also, unlike other systems, \fluto\ allows for establishing optimality and address the strict stoichiometric completion problem without approximation.

\fluto\ takes advantage of the hybrid reasoning capacities of the ASP system \clingo{}
for extending logic programs with linear constraints over reals.
This provides us with a practically relevant application scenario for evaluating this hybrid form of ASP.
To us, the most surprising empirical result was the observation that domain-specific heuristic allow for boosting unsatisfiable core based
optimization.
So far, such heuristics have only been known to improve satisfiability-oriented reasoning modes, and usually hampered unsatisfiability-oriented ones
(cf.~\citep{gekakarosc15a}).


\paragraph{Acknowledgments}

This work was partially funded by DFG grant SCHA~550/9 and~11 and benefited from the support of the French Government via the National Research Agency investment expenditure program IDEALG ANR-10-BTBR-04.


\bibliographystyle{acmtrans}
\bibliography{lit,procs,akku,bibioinfo} 
\newpage
\appendix

\section{Factual representation of example metabolic network}
\label{sec:appendix}

The factual representation of the metabolic network in Fig.~\ref{gra:toy} is given in Listing~\ref{lst:instance}.

\begin{lstlisting}[numbers=left,numberblanklines=false,basicstyle=\ttfamily\footnotesize,caption={Example instance of metabolic network},label=lst:instance]
metabolite("S1",s). metabolite("S2",s). metabolite("S3",s).
metabolite("a",t).  metabolite("b",d).  metabolite("c",t).
metabolite("d",d).  metabolite("e",d).  metabolite("f",d).
metabolite("g",r).

reaction("R_importS1",s).     reaction("R_importS2",s).
reversible("R_importS1").     reversible("R_importS2").
prd("S1","1","R_importS1",s). prd("S2","1","R_importS2",s).
reaction("R_exportF",d).      rct("f","1","R_exportF",d).

reaction("R0",d).      reaction("R1",d).
rct("b","1","R0",d).   rct("S3","1","R1",d).
prd("S3","1","R0",d).  prd("b","1","R1",d).

reaction("R2",d).      reaction("R3",d).
rct("c","1","R2",d).   rct("d","1","R3",d).
rct("S2","1","R2",d).  prd("e","1","R3",d).
prd("d","1","R2",d).

reaction("R4",d).      reaction("R9",r).
rct("e","1","R4",d).   rct("g","1","R9",r).
prd("c","2","R4",d).   prd("f","1","R9",r).

reaction("R5",t).      reaction("R6",r).
rct("a","1","R5",t).   rct("S1","1","R6",r).
rct("c","1","R5",t).   prd("a","1","R6",r).
prd("f","1","R5",t).   prd("g","1","R6",r).

reaction("R7",r).      reaction("R8",r).
rct("S3","1","R7",r).  rct("b","1","R8",r).
prd("e","1","R7",r).   prd("e","1","R8",r).

objective(R,T) :- reaction(R,T), T!=t.
objective(R,t) :- reaction(R,t).

bounds(R,"0","99999") :- reaction(R,_), not reversible(R).
bounds(R,"-99999","99999") :- reaction(R,_), reversible(R).
\end{lstlisting}

Note that in lines 33 to 37 of Listing~\ref{lst:instance}, 
the values of \texttt{objective} and \texttt{bounds} are set globally,
but they may be arbitrary in general.


\end{document}